\documentclass[aps,prd,a4paper,longbibliography,reprint,superscriptaddress,floatfix,nofootinbib,amsmath,amssymb]{revtex4-1}
\usepackage[latin2]{inputenc}
\usepackage{dsfont}
\usepackage{subfigure}
\usepackage{mathrsfs}
\usepackage{slashed}
\usepackage{array}
\usepackage{colordvi}
\usepackage{rotfloat}
\usepackage{rotating}
\usepackage{color}
\usepackage[    bookmarks,
                 bookmarksopen = true,
                 bookmarksnumbered = true,
                 linktocpage,
                 colorlinks = true,
                 linkcolor = blue,
                 urlcolor  = blue,
                 citecolor = blue,
                 anchorcolor = green,
                 hyperindex = true,
                 hyperfigures]
		{hyperref}
\usepackage{cmap}
\newcommand{\tr}{\mathrm{tr}\,}

\begin{document}

\title{Non-perturbative determination of improvement coefficients using coordinate space correlators in \boldmath$N_f=2+1$ lattice QCD} 

\author{Piotr Korcyl}
\email[]{piotr.korcyl@ur.de}
\affiliation{Institut f\"ur Theoretische Physik, Universit\"at Regensburg, D-93040 Regensburg, Germany}
\author{Gunnar S. Bali}
\email[]{gunnar.bali@ur.de}
\affiliation{Institut f\"ur Theoretische Physik, Universit\"at Regensburg, D-93040 Regensburg, Germany}
\date{\today}
\begin{abstract}
We determine quark mass dependent order $a$ improvement terms of the form
$b_Jam$ for non-singlet scalar, pseudoscalar, vector and axialvector currents
using correlators in coordinate space on
a set of CLS ensembles. These have been generated employing
non-perturbatively improved Wilson Fermions and the tree-level
L\"uscher-Weisz gauge action at $\beta=3.4, 3.46, 3.55$ and $3.7$,
corresponding to lattice spacings ranging from
$a\approx 0.085\,\textmd{fm}$ down to $0.05\,\textmd{fm}$.
In the $N_f=2+1$ flavour theory two types of improvement coefficients
exist: $b_J$, proportional to non-singlet quark mass combinations, and
$\bar{b}_J$ (or $\tilde{b}_J$), proportional to the trace of the
quark mass matrix.
Combining our non-perturbative determinations with
perturbative results, we quote Pad\'e approximants parameterizing
the $b_J$ improvement coefficients within the above window of lattice
spacings. We also give preliminary results for $\tilde{b}_J$ at $\beta=3.4$.
\end{abstract}

\maketitle

\section{Introduction}
Lattice simulations of quantum chromodynamics (Lattice QCD)
have become an indispensable tool in
particle and hadron physics phenomenology. By discretizing
a quantum field theory on a lattice with a spacing $a>0$,
ultraviolet divergences are regularized. At the same time
this enables the numerical simulation of QCD, including its
non-perturbative dynamics. In principle such simulations need to be
performed for different values of the lattice spacing, in order to
remove the regulator by taking the continuum
limit, $a\rightarrow 0$. In QCD this limit is approached as a polynomial
in $a$, modulated by logarithmic corrections.

Obviously, many possible discretizations of
the quark (and gluon) parts of the action exist. Staggered
quarks suffer from conceptional problems, unless 
all Fermions come in mass degenerate groups of four flavours.
Also combining the flavour and spin degrees of freedom
complicates operator mixing and the analysis of two- and
three-point Green functions. Domain wall and overlap actions
have the most desirable theoretical properties as
even at a non-vanishing value of the lattice spacing 
these possess an (almost) exact chiral symmetry in the massless limit.
In contrast, using Wilson Fermions,
chiral symmetry only becomes restored in the continuum limit,
and also an additive mass renormalization is encountered.
Wilson Fermions, however, are computationally much less
expensive to simulate and therefore offer the possibility of
obtaining results at several values of the lattice spacing, enabling
a controlled continuum limit extrapolation.

Unlike other Fermion discretizations, where leading lattice artefacts
are of order $a^2$, for naive Wilson Fermions these are
linear in $a$. Such terms can, however, be removed
non-perturbatively~\cite{Luscher:1996sc,Luscher:1996ug},
Symanzik improving~\cite{Symanzik:1983dc} the action and the
local operators of interest. Recently, within the
Coordinated Lattice Simulations (CLS) effort~\cite{Bruno:2014jqa},
we embarked on a large scale simulation programme, employing $N_f=2+1$
flavours of order $a$ improved
Wilson-Sheikholeslami-Wohlert~\cite{Sheikholeslami:1985ij} (clover) Fermions
and the tree-level improved L\"uscher-Weisz
gauge action~\cite{Weisz:1982zw,Luscher:1984xn}.
CLS use open boundary conditions in time~\cite{Luscher:2011kk},
thereby increasing the mobility of
topological charges and enabling us to maintain ergodicity
at finer lattice spacings than had been possible previously.
For details on the action, ensembles and parameter values,
see Ref.~\cite{Bruno:2014jqa}.

As the cost of simulations increases with a large inverse
power of the lattice spacing, we aim at not only order $a$ improving
the action but also all operators that will appear in matrix elements
of interest. It is important to remove such contributions, that are
linear in $a$, non-perturbatively
since terms of order $g^{2\nu}a$, where $g$ denotes the gauge coupling,
will survive a $(\nu-1)$-loop perturbative subtraction.
As $g^2$ varies only slowly with $a$, close to the
continuum limit any $g^{2\nu}a$ term will dominate over $a^2$ terms.
The non-perturbative improvement of the action and of the massless axial
current was carried out in Refs.~\cite{Bulava:2013cta,Bulava:2015bxa}.
In addition to such ``$c_J$'' improvement terms that persist in the
massless limit, in the massive case additional $b_J$ and $\bar{b}_J$
coefficients are encountered for a current $J$,
for definitions, see, e.g., Ref.~\cite{Bhattacharya:2005rb}.
Existing results as of 2006 are reviewed in Ref.~\cite{Sommer:2006sj}
and, more recently, for $N_f=2$ clover quarks on Wilson glue,
the combinations $b_A-b_P$ and $b_S=-2b_m$ were determined
non-perturbatively in Ref.~\cite{Fritzsch:2010aw}.

Here we introduce a variant of the coordinate space method
that was originally proposed in
Ref.~\cite{Martinelli:1997zc}. This will allow us to determine the
scalar, pseudoscalar, vector and axial
$b_J$ coefficients, accompanying both flavour-singlet and non-singlet
quark mass combinations,
with very limited computational effort. This is then successfully
applied to the CLS ensembles described above.

This article is organized as follows.
In Sec.~\ref{sec. description} we describe the general 
approach and define the observables that will be studied.
Next, in Sec.~\ref{sec. tree-level} we analyse these observables at 
tree-level in lattice and continuum perturbation theory, including
the leading non-perturbative effects, that are expected from the
operator product expansion. This will allow us to improve the
observables, to estimate the size
of cut-off effects and to select the optimal 
set of separations at which the correlation functions are
evaluated in the non-perturbative study. 
Then in Sec.~\ref{sec. systematics} we discuss systematic errors
of our approach, addressing finite volume effects
and estimating contributions of non-perturbative condensates.
Finally, in Sec.~\ref{sec. results} we present results for all
order $am$ coefficients. In Sec.~\ref{sec. conclusions}
we conclude and present an outlook.

\section{Description of the method}
\label{sec. description}

We generalize the method of Ref.~\cite{Martinelli:1997zc}
to the situation of $N_f=2+1$ non-degenerate quark mass flavours.
For increased precision, we perturbatively subtract the leading
order lattice artefacts. Furthermore, we employ the operator product
expansion (OPE), enabling us to quantitatively
describe medium distance corrections.

We will assume improved Wilson quarks and --- as we aim at
order $a$ improvement --- we will consequently drop all
terms of order $a^2$. We remark that different prescriptions
of obtaining improvement coefficients will in general give results
that differ by such higher order corrections.

We denote quark mass averages
following Refs.~\cite{Bali:2016umi,Bhattacharya:2005rb} as
\begin{equation}
m_{jk} = \frac{1}{2}(m_j + m_k)\,,
\end{equation}
where
\begin{equation}
m_j = \frac{1}{2a} \left( \frac{1}{\kappa_j} - \frac{1}{\kappa_{\textrm{crit}}} \right).
\end{equation}
The critical hopping parameter value $\kappa_{\textrm{crit}}$ is defined
as the point where the axial Ward identity (AWI) quark mass vanishes
in the theory with $N_f=3$ mass degenerate quark flavours.

We will label the mass of the two degenerate quark flavours as
$m_1=m_2=m_{\ell}$ and the mass of the remaining (strange) quark 
as $m_3=m_s$. The mass dependence of physical observables can be
parameterized in terms of the average quark mass
\begin{equation}
\overline{m} = \frac{1}{3}\left(m_s + 2 m_{\ell}\right)\,,
\end{equation}
and the light quark mass $m_{12}$ or, equivalently, the average
of the strange and light quark masses $m_{13}$:\footnote{It is
not necessary to differentiate between lattice and renormalized
quark masses in the present context.}
if $\overline{m}$ and either $m_{12}$ or $m_{13}$ are known,
$m_{\ell}$ and $m_s$ are fixed. Most ensembles have been generated
following the strategy of the QCDSF Collaboration~\cite{Bietenholz:2010jr},
keeping $\overline{m}$ constant. This is supplemented by further
ensembles at an (approximately) fixed value of the
renormalized strange quark mass, as well
as along the symmetric line $m_{\ell}=m_s$ \cite{Bali:2016umi}.
 
\subsection{General considerations and definitions}
We define connected
Euclidean current-current correlation functions in a continuum
renormalization scheme $R$, e.g., $R=\overline{\mathrm{MS}}$,
at a scale $\mu$:
\begin{equation}
\label{eq:green}
G_{J^{(jk)}}^R(x,m_{\ell},m_s;\mu)=\left\langle\Omega\left|T\,J^{(jk)}(x)\overline{J}^{(jk)}(0)\right|\Omega\right\rangle^R.
\end{equation}
$T$ denotes the time ordering operator, which we shall omit below as
path integral expectation values are automatically time ordered.
$|\Omega\rangle$ is the vacuum state and $J\in\{S,P,V_{\mu},A_{\mu}\}$.
The current is defined as
\begin{equation}
J^{(jk)}=\overline{\psi}_j\Gamma_J\psi_k\,,\quad
\overline{J}^{(jk)}=\overline{\psi}_k\Gamma_J^{\dagger}\psi_j\,,
\label{eq:conve}
\end{equation}
with $\Gamma_J\in\{\mathds{1},\gamma_5,\gamma_{\mu},\gamma_{\mu}\gamma_5\}$.
$\psi_j$ destroys a quark of flavour $j\in\{1,2,3\}$ and
$x$ is a four-distance vector in coordinate space.
As here we will only consider flavour non-singlet currents,
we always assume $j\neq k$.

The above correlation function differs from that of the massless case
by mass dependent terms~\cite{Reinders:1981sy,Reinders:1984sr,Jamin:1992se},
\begin{align}
\label{eq.massless}
&G_{J^{(jk)}}^R(x,m_{\ell},m_s;\mu)=G_{J^{(jk)}}^R(x,0,0;\mu)\\\nonumber
\times&
\left[1+\mathcal{O}\left(m^2x^2,m^2\langle FF\rangle x^6,m\langle \overline{\psi}\psi\rangle x^4,m\langle\overline{\psi}\sigma F\psi\rangle x^6\right)\right]\,,
\end{align}
where at each order in $m$ we
only display the dominant type of term. Note that only even powers of $x$
can appear above. Regarding the non-perturbative correction terms,
the light quark condensate (in the $\overline{\mathrm{MS}}$ scheme
at the scale $\mu=2\,\textmd{GeV}$) reads
$\langle\overline{\psi}\psi\rangle=-\Sigma_0=-[274(3)\,\textmd{MeV}]^3$~\cite{Aoki:2016frl}.
Recently, the renormalization group invariant
non-perturbative gluon condensate $\langle FF\rangle$ was
determined from a high order perturbative expansion
in $\textmd{SU}(3)$ gauge theory~\cite{Bali:2014sja}, with the result
$\langle FF\rangle \sim (530\,\textmd{MeV})^4$ being larger than
the original estimate
$\langle FF\rangle \sim (330\,\textmd{MeV})^4$~\cite{Shifman:1978bx}.
Unlike the quark condensate, this object is ill-defined in principle
and the uncertainty
of its definition was determined to be similar in
magnitude to its size~\cite{Bali:2014fea,Bali:2015cxa}. The Wilson coefficient
accompanying the
$m^2\langle FF\rangle$ term
reads at leading
order $1/12$ for $S$ and $P$
and $1/6$ for $V$ and $A$~\cite{Reinders:1984sr,Jamin:1992se}, and
$\langle FF\rangle/6\sim (340\,\textmd{MeV})^4$, even if we assume
the higher value~\cite{Bali:2014sja} for $\langle FF\rangle$.
The mixed condensate~\cite{Narison:1983kn}
is usually estimated to be
$|\langle\overline{\psi}\sigma F\psi\rangle|\sim 0.8\, \textmd{GeV}^2|\langle\overline{\psi}\psi\rangle|\sim(430\,\textmd{MeV})^5$~\cite{Narison:1988ep}.
To leading order the Wilson coefficient accompanying
this condensate reads $m/2$ for $S$ and $P$ but vanishes
for $A$ and $V$~\cite{Jamin:1992se}.
We conclude that all mass dependent condensate contributions
are bound by a respective power of
a scale $\Lambda \approx 400\,\textmd{MeV}$.
Then, in the limit
\begin{equation}
x^{-2}\gg \max\left\{m_s^{1/3}\Lambda^{5/3}, m_s^{1/2} \Lambda^{3/2}, m_s^{2/3}\Lambda^{4/3}, m_s^2\right\},
\end{equation}
the higher order terms in Eq.~\eqref{eq.massless} can be neglected.
Assuming $\Lambda>m_s\ge m_{\ell}$, we arrive at 
the condition $x^2\ll 1/\Lambda^2$, i.e.\ $|x|$ needs to be much
smaller than $0.5\,\textmd{fm}$ to permit neglecting
mass dependent terms on the continuum side.
In Sec.~\ref{sec. tree-level} below we will carry out a detailed
analysis of the leading mass dependent corrections to Eq.~\eqref{eq.massless}.

The continuum Green function $G^R$ above can be related to
the corresponding Green function $G$ obtained in the lattice scheme
at a lattice spacing $a=a(g^2)$ as follows:
\begin{align}
\label{eq:ren}
&G_{J^{(jk)}}^R(x,m_{\ell},m_s;\mu)=\left(Z_J^R\right)^2\!(\tilde{g}^2,a\mu)
\\\nonumber
&\times\left(1+2b_Jam_{jk}+6\bar{b}_Ja\overline{m}\right)
G_{J^{(jk),I}}(n,am_{jk},a\overline{m};g^2)\,,
\end{align}
where $x=na$, $n_{\mu}\in\mathbb{Z}$ so that $n^2=n_{\mu}n_{\mu}$ is
integer-valued and~\cite{Luscher:1996sc}
$\tilde{g}^2=(1+b_ga\overline{m})g^2$ is
the order $a$ improved value of the bare lattice
coupling $g^2=6/\beta$. Not only $Z_J^R$ but also $b_J$ and $\bar{b}_J$ will
depend on $\tilde{g}^2$ rather than on $g^2$, however,
we can drop order $a$ corrections to
order $a$ improvement coefficients and substitute
$b_J(\tilde{g}^2)$ and $\bar{b}_J(\tilde{g}^2)$ by
$b_J(g^2)$ and $\bar{b}_J(g^2)$.

Expanding $Z_J^R$ around $g^2$ gives~\cite{Bali:2016umi}
\begin{widetext}
\begin{align}
\nonumber
Z_J^R\left[\tilde{g}^2,a(\tilde{g}^2)\mu\right]&=
Z_J^R\left[g^2,a(g^2)\mu)\right]\left[1+
\left(\frac{\partial \ln Z_J^R(g^2,a\mu)}{\partial g^2}+
\frac{\partial\ln Z_J^R(g^2,a\mu)}{\partial\ln a}\frac{\mathrm{d}\ln a(g^2)}{\mathrm{d}{g^2}}\right)g^2b_ga\overline{m}+\ldots\right]\\
&=
Z_J^R(g^2,a(g^2)\mu)\left\{1+\left[\frac{\partial\ln{Z_J^R(g^2,a\mu)}}{\partial g^2}-\frac{\gamma_J(g^2)}{4\pi\beta(g^2)}\right] b_gg^2a\overline{m}\right\}\,,
\end{align}
\end{widetext}
where
\begin{equation}
\beta(g^2)=-\frac{1}{4\pi}\frac{\mathrm{d}g^2}{\mathrm{d}\ln a}=
-\frac{g^2}{2\pi}\left[\beta_0\frac{g^2}{16\pi^2}
+\cdots\right]
\end{equation}
is the QCD $\beta$-function in the normalization convention
$\beta_0=11-\frac23N_f$. The anomalous dimension of the current 
$J$ reads,
\begin{equation}
\gamma_J(g^2)=\frac{\mathrm{d}\ln Z_J}{\mathrm{d}\ln a}\,,
\end{equation}
and is trivial for $A_{\mu}$ and $V_{\mu}$.
We can eliminate $b_g$ by redefining
\begin{align}
\label{eq:redef}
\tilde{b}_J(g^2)&=
\bar{b}_J(g^2)\\\nonumber&+\frac{b_g(g^2)}{N_f}
\left[\frac{\partial\ln Z_J^R(g^2,a\mu)}{\partial g^2}
-\frac{\gamma_J(g^2)}{4\pi\beta(g^2)}\right]g^2\,.
\end{align}
Both $\bar{b}_J$ and
$\tilde{b}_J$ are of $\mathcal{O}(g^4)$ in perturbation theory.
$Z_J^R$~\cite{Taniguchi:1998pf,Constantinou:2014fka} and
$b_g=0.012000(2)N_fg^2$~\cite{Luscher:1996sc} are known
to $\mathcal{O}(g^2)$ and,
therefore, the difference between $\bar{b}_J$ and $\tilde{b}_J$ is available to
$\mathcal{O}(g^4)$. However, the $\bar{b}_J$ coefficients are at present
not available to this first non-trivial order. So the only thing we know
is that $\tilde{b}_J=\mathcal{O}(g^4)$.
Below we will also estimate these coefficients
non-perturbatively. For practical purposes determining $\tilde{b}_J$ is sufficient as
a knowledge of $\bar{b}_J$ is usually not required.

With the above redefinitions Eq.~\eqref{eq:ren} reads:
\begin{align}
\label{eq:ren2}
&G_{J^{(jk)}}^R(x,m_{\ell},m_s;\mu)=\left[Z_J^R(g^2,a\mu)\right]^2
\\\nonumber
&\times\left(1+2b_Jam_{jk}+6\tilde{b}_Ja\overline{m}\right)
G_{J^{(jk),I}}(n,am_{jk},a\overline{m};g^2)\,.
\end{align}

The superscript $I$ of $J^{(jk),I}$ on the
right hand sides of Eqs.~\eqref{eq:ren} and \eqref{eq:ren2}
refers to order $a$ improved lattice currents:
$S^{(jk),I}=S^{(jk)}$, $P^{(jk),I}=P^{(jk)}$,
$V_{\mu}^{(jk),I}=V_{\mu}^{(jk)}+iac_V\partial_{\nu}
T_{\mu\nu}^{(jk)}$,
$A_{\mu}^{(jk),I}=A_{\mu}^{(jk)}+
ac_A\partial_{\mu}P^{(jk)}$, where $T_{\mu\nu}^{(jk)}=\overline{\psi}_j\sigma_{\mu\nu}
\psi_k$, $\sigma_{\mu\nu}=\frac{i}{2}[\gamma_{\mu},\gamma_{\nu}]$ and
$\partial_{\mu}$ denotes the symmetric next neighbour lattice derivative:
$\partial_{\mu}f(x)=[f(x+a\hat{\mu})-f(x-a\hat{\mu})]/2a$.

Here we will consider the following correlators:
\begin{align}
G_{S^{(jk)}}(x) &= \left\langle S^{(jk)}(x)
\overline{S}^{(jk)}(0) \right\rangle
= G_{S^{(jk),I}}(x)\,,
 \label{eq corr s 12} \\
G_{P^{(jk)}}(x) &= 
\left\langle P^{(jk)}(x)
\overline{P}^{(jk)}(0) \right\rangle
= G_{P^{(jk),I}}(x)\,,
\label{eq corr p 12} \\
\label{eq corr v 12} G_{V^{(jk)}}(x) &= \frac{1}{4} \sum_{\mu}
\left\langle V_{\mu}^{(jk)}(x)
\overline{V}_{\mu}^{(jk)}(0) \right\rangle\\\nonumber
&= G_{V^{(jk),I}}(x)\left[1+\mathcal{O}(a)\right]\,,
\\
\label{eq corr a 12}
G_{A^{(jk)}}(x) &= \frac{1}{4} \sum_{\mu}
\left\langle A_{\mu}^{(jk)}(x)
\overline{A}_{\mu}^{(jk)}(0) \right\rangle\\\nonumber
&= G_{A^{(jk),I}}(x)\left[1+\mathcal{O}(a)\right]\,,
\end{align}
where we suppressed the arguments $m_{jk}$ and $\overline{m}$.
We remark that $c_A$ is known
non-perturbatively~\cite{Bulava:2015bxa} for
the action in use. In principle it can also be determined
with coordinate space methods~\cite{Martinelli:1997zc}, tuning
$\sum_{\mu}\langle [A_{\mu}^I(x_1)-A_{\mu}^I(x_2)]\overline{P}(0)\rangle=0$
for two Euclidean distances $x_1\neq x_2$ with $x_1^2=x_2^2$.
In this study we employ unimproved currents since mass independent
order $a$ corrections cancel from the ratios that we will consider.

\subsection{Description of the method}
For the moment being we assume $x^2$ to be much smaller than $\Lambda^{-2}$.
Then
the continuum Green function for massive quark currents is well approximated
by the massless one and we can write
\begin{align}
\label{eq:greenstuff}
&G_{J^{(jk)}}^R(x,0,0;\mu)\stackrel{x^2\ll \Lambda^{-2}}{\approx}
\left[Z_J^R(g^2,a\mu)\right]^2\\\nonumber
&\left(1+2b_Jam_{jk}+6\tilde{b}_Ja\overline{m}\right)
G_{J^{(jk)}}(n,am_{jk},a\overline{m};g^2)\,.
\end{align}
Above we omited mass independent order $a$ corrections,
which exist for $J=V$ and $J=A$,
see Eqs.~\eqref{eq corr v 12}--\eqref{eq corr a 12},
since these will cancel from the ratios that we are going to consider.

Obviously, in the massless limit, the ratio of two continuum Green functions
$G_{J^{(jk)}}^R(x;\mu) \equiv G_{J^{(jk)}}^R(x,0,0;\mu)$
for the same current $J$ but different or the same $jk$ flavour combinations
cancels. We have discussed above that mass dependent corrections
to this continuum ratio are proportional to orders of $x^2$.
Thus, we obtain,
\begin{widetext}
\begin{equation}
\label{eq:ratio}
\frac{G_{J^{(jk)}}\left(n,am^{(\rho)}_{jk},a\overline{m}^{(\rho)};g^2\right)}
{G_{J^{(rs)}}\left(n,am^{(\sigma)}_{rs},a\overline{m}^{(\sigma)};g^2\right)}
=
1+2b_Ja\left(m^{(\sigma)}_{rs}-m^{(\rho)}_{jk}\right)+6\tilde{b}_Ja\left(\overline{m}^{(\sigma)}-\overline{m}^{(\rho)}\right)+\mathcal{O}\left(a^2,x^2\right),
\end{equation}
\end{widetext}
where $\rho$ and $\sigma$ refer to different simulation points
in the quark mass plane at a fixed value of the coupling $g^2$ and
the indices $j,k,r,s\in\{1,2,3\}$ refer to the three flavours.
While $j\neq k$ and $r\neq s$, $\rho=\sigma$ is allowed.
Combining results for different pairs of quark masses therefore
enables us to determine the $b_J$ and $\tilde{b}_J$ coefficients.
Note that as only quark mass differences appear above, no knowledge
of $\kappa_{\mathrm{crit}}$ is required. This can only become
relevant for the improvement of flavour singlet currents.

The leading $a$ dependent correction terms can be of the
types $a^2m^2$, $a^2m\Lambda$ (for $J=V$ and $J=A$),
$a^3m\Lambda^2$ and $a^3m/x^2=am/n^2$. In contrast,
the $x^2=(na)^2$, $x^4$ etc.\ corrections are no
lattice artefacts but have well-defined continuum limits.
This means that the determination
of the improvement coefficients becomes possible for
$x^2\ll 1/\Lambda^2$ but the precision is limited by the size of $1/n^2
=a^2/x^2$, resulting in the window $a^2< x^2\ll\Lambda^{-2}$.

We remark that unlike in determinations of the renormalization
constants $Z_J$~\cite{Cichy:2012is,Gimenez:2004me,Tomii:2016xiv},
as long as $x^2$ is within the above
window, no knowledge on the functional form of $G_J(x)$ is required
to extract $b_J$ and $\tilde{b}_J$. Moreover, short-distance
lattice artefacts are much reduced within the above ratio.
Nevertheless, in Sec.~\ref{sec. tree-level}
below we will correct for the leading order
lattice artefacts as well as for the leading $x^2$ 
and $x^4$ correction terms to Eq.~\eqref{eq:ratio},
to broaden the window of distances where the
method described can be applied.

\subsection{Observable for the \boldmath$b_{J}$ coefficients}
\label{sec. method}
We consider a ratio $R_J(x,m_{12},m_{13})$ of two correlators evaluated on
a single ensemble,
i.e.\ we employ Eq.~\eqref{eq:ratio} with $\rho = \sigma$. 
As $\overline{m}$ is fixed, $R_J$ only depends
on $m_s-m_{\ell}=2(m_{13}-m_{12})$:
\begin{align}
R_J(x, m_s-m_{\ell}) &\equiv \frac{G_{J^{(12)}}\left(n,am^{(\rho)}_{12},a\overline{m}^{(\rho)};g^2\right)}
{G_{J^{(13)}}\left(n,am^{(\rho)}_{13},a\overline{m}^{(\rho)};g^2\right)}\nonumber\\
 &= 
1 + 2 b_J a \left(m_{13}^{(\rho)} - m_{12}^{(\rho)}\right)\nonumber\\
&=1+\frac{1}{2}b_J\left(\frac{1}{\kappa_s}-\frac{1}{\kappa_{\ell}}\right)\,.
\label{eq:rrat}
\end{align}
Hence, $R_J-1$ is directly proportional to $b_J$, with a
known prefactor that is independent of $\kappa_{\mathrm{crit}}$.
Therefore, to determine $b_J$, a single measurement at a simulation point
with $\kappa_{\ell}\neq\kappa_s$ is sufficient.

\begin{figure}
\includegraphics[width=0.475\textwidth]{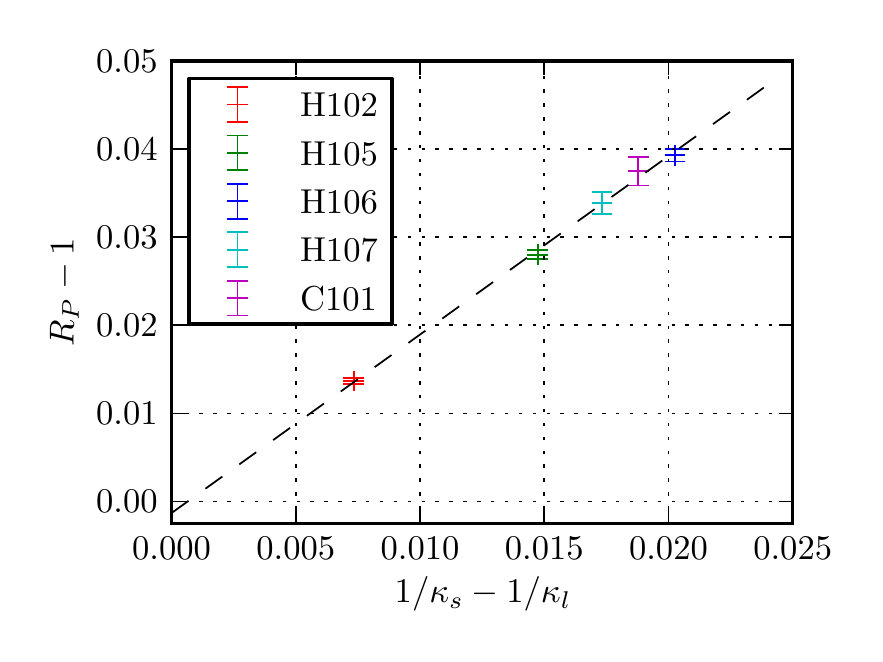}
\caption{Dependence of the ratio $R_P-1$ at the separation
$x=(0,1,1,1)a$ as a function of the inverse hopping parameter
difference $1/\kappa_s-1/\kappa_{\ell}$,
see Eq.~\eqref{eq:rrat}. For an ensemble list,
see Table~\ref{ensemble table}.\label{fig. demo}}
\end{figure}

In Fig.~\ref{fig. demo} we demonstrate this for $J=P$, by showing
$R_P(x, m_{s} - m_{\ell})-1$ at a fixed separation $x=(0,1,1,1)a$ and value
of the lattice spacing $a\approx 0.085\,\textmd{fm}$ ($\beta=3.4$)
as a function of $1/\kappa_s - 1/\kappa_{\ell}$.
This is carried out on different ensembles.
As expected, the data lie on a straight line whose slope is proportional
to $b_P$. The intercept is at the origin as there are no
mass independent order $a$ effects in the current setting.
The fact that a linear fit is consistent with this intercept
demonstrates that the next-to-leading order lattice artefacts, that are
proportional to $m^2a^2$, are small at our quark mass values.
The point $x$ shown in this example appears to be well suited for the
extraction of $b_P$. Below
we will provide criteria to optimize this choice.

\subsection{Observable for the \boldmath$\tilde{b}_{J}$ coefficients}
In contrast to the improvement coefficients $b_J$ accompanying non-singlet
mass combinations, the $\tilde{b}_J$ coefficients can only be determined
varying the average quark mass $\overline{m}$. 
Again, we start from the ratio of correlation functions
Eq.~\eqref{eq:ratio}, where this time $\rho \neq \sigma$ is necessary,
i.e.\ information from at least two ensembles needs to be combined.
The main set of CLS simulations~\cite{Bruno:2014jqa} is obtained
along a trajectory of constant $\overline{m}$, where no sensitivity
to $\tilde{b}_J$ exists. However, we have points on two additional mass plane
trajectories at our disposal (for details, see Ref.~\cite{Bali:2016umi}),
one along which only the light quark
mass is varied while the AWI strange quark mass is kept constant
and one line, along which $\kappa_{\ell}=\kappa_s=\kappa^{(\rho)}$, i.e.\
$\overline{m}^{(\rho)}=m_{12}^{(\rho)}=m_{13}^{(\rho)}$. A coefficient
$\tilde{b}_J$ can most
easily be obtained along this ``symmetric'' trajectory,
once $b_J$ is known,
as in this case
\begin{align}
\widetilde{R}_J(x, \overline{m}^{(\sigma)}-\overline{m}^{(\rho)}) &\equiv \frac{G_{J^{(12)}}
\left(n,a\overline{m}^{(\rho)},a\overline{m}^{(\rho)};g^2\right)}
{G_{J^{(12)}}\left(n,a\overline{m}^{(\sigma)},a\overline{m}^{(\sigma)};g^2\right)}
\nonumber\\
&=1+(2b_J+6\tilde{b}_J)a
\left(\overline{m}^{(\sigma)}-\overline{m}^{(\rho)}\right)\nonumber \\
&=
1+\frac{1}{2}(b_J+3\tilde{b}_J)\left(\frac{1}{\kappa^{(\sigma)}}-\frac{1}{\kappa^{(\rho)}}\right).
\label{eq:rrat2}
\end{align}
Note that
again no knowledge of $\kappa_{\mathrm{crit}}$ is required.
It is also possible to employ a pair of ensembles that differ
in their $\kappa_s$ value but share a similar $\kappa_{\ell}$ value,
eliminating the dependence on $b_J$ altogether when only light quark
correlation functions are considered.

\begin{figure}
\includegraphics[width=0.475\textwidth]{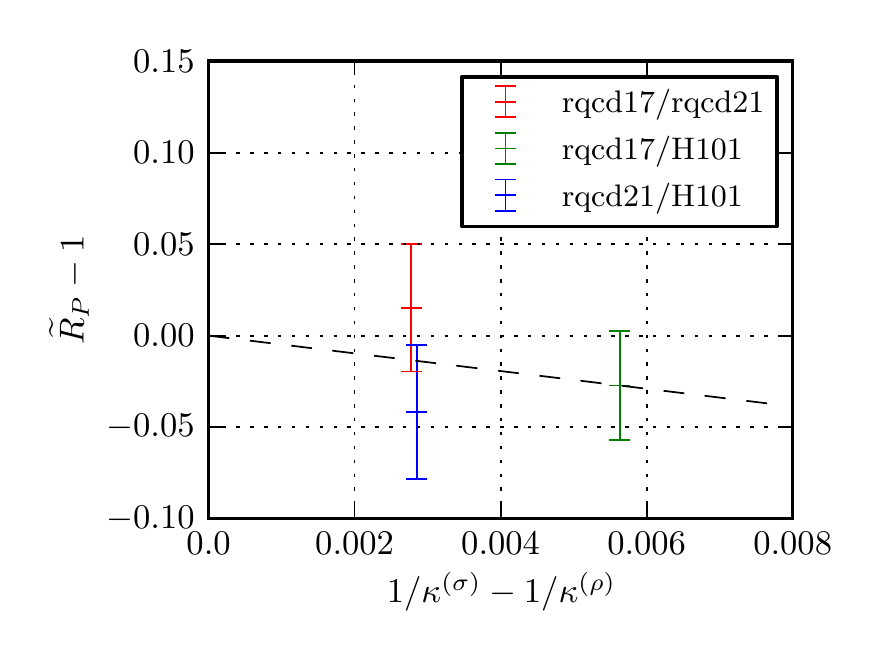}
\caption{Dependence of the ratio $\widetilde{R}_P-1$ at the
separation $x=(0,1,2,2)a$ as a function of the inverse hopping
parameter difference
$1/\kappa^{(\sigma)}-1/\kappa^{(\rho)}$.\label{fig. demo bar}}
\end{figure}

We demonstrate how $\tilde{b}_P$ can be extracted from the slope of
$\widetilde{R}_J$ in
Fig.~\ref{fig. demo bar}:
Using a set of three $m_{\ell}=m_s$ ensembles at $\beta=3.4$, we evaluate the
ratio of correlators for all three possible pairs of ensembles.
Note that the statistical errors are much larger than in the case of
$b_P$, mostly because the numerator and denominator of Eq.~\eqref{eq:rrat2}
are uncorrelated.

\section{Behaviour at short and long distances}
\label{sec. tree-level}
In this section we discuss corrections to the ratios $R_J(x, \Delta m)$
and $\widetilde{R}(x,\Delta \overline{m})$, see
Eqs.~\eqref{eq:rrat} and \eqref{eq:rrat2}, at large and short distances
and define improved observables.
We employ Euclidean spacetime conventions throughout.
\subsection{Continuum expectation}
The components of the propagator $S_F(x)\equiv S_F(x,0)$
for a quark $\psi^i_{\alpha}$ propagating
from a four-position $0$ to $x$ are given as,
\begin{equation}
S_{F\alpha\beta}^{ij}(x)=\psi^i_{\alpha}(x)\overline{\psi}_{\beta}^j(0)
=-\overline{\psi}_{\beta}^j(0)\psi^i_{\alpha}(x)\,,
\end{equation}
where $\alpha,\beta$ denote spinor and $i,j$ colour indices.
This receives perturbative and non-perturbative contributions.
The massive free case propagator reads (see, e.g., Ref.~\cite{Chu:1993cn}):
\begin{align}\nonumber
S_F(x) &= \frac{1}{(2\pi)^2} \left[
\gamma_{\mu} x_{\mu} \frac{m^2}{x^2} K_2(m|x|) + \frac{m^2}{|x|} K_1(m |x|) \right] \\
&= \frac{1}{2\pi^2}\frac{\gamma_{\mu}x_{\mu}}{x^4}\left(1-\frac{m^2x^2}{4}\right)
+\frac{m}{4\pi^2}\frac{1}{x^2}+\cdots,
\end{align}
where $K_1(z)$ and $K_2(z)$ are modified Bessel functions of the second kind.
The leading non-perturbative contributions can be obtained,
expanding
\begin{equation}
S_{F\alpha\beta}^{ij}(x)=-\overline{\psi}_{\beta}^j(0)\psi^i_{\alpha}(0)
-x_{\mu}\overline{\psi}_{\beta}^j(0)[D_{\mu}\psi_{\alpha}]^i(0)+\cdots.
\end{equation}
The colour, spinor and Lorentz structure then implies that
\begin{align}
\langle \overline{\psi}_{\beta}^j\psi^i_{\alpha}\rangle&=b\delta^{ij}\delta_{\alpha\beta}\,,\\
\langle \overline{\psi}_{\beta}^j[D_{\mu}\psi_{\alpha}]^i\rangle
&=c\delta^{ij}\left(\gamma_{\mu}\right)_{\alpha\beta},
\end{align}
where the constants $b$ and $c$ are easily determined:
\begin{align}
\langle \overline{\psi}\psi\rangle&=\sum_{i,\alpha}
\langle \overline{\psi}_{\alpha}^i\psi_{\alpha}^i\rangle=4Nb\,,\\\nonumber
m\langle \overline{\psi}\psi\rangle&=
-\gamma_{\mu}\langle \overline{\psi}D_{\mu}\psi\rangle
\\&=-\sum_{i,\alpha,\beta,\mu}
\left(\gamma_{\mu}\right)_{\alpha\beta}\langle \overline{\psi}_{\beta}^i[D_{\mu}\psi_{\alpha}]^i\rangle\nonumber\\
&=-cN\sum_{\mu}\tr\gamma_{\mu}\gamma_{\mu}=-16Nc\,.
\end{align}
Above we made use of the equations of motion and $N=3$ is the number
of colours.

Collecting our results gives
\begin{align}
S_F^{ij}(x)&=f(x)\delta^{ij}\mathds{1}+g_{\mu}(x)\gamma_{\mu}\delta^{ij}+\cdots,\\
f(x)&=\frac{m}{4\pi^2}\frac{1}{x^2}-\frac{1}{4N}\langle\overline{\psi}\psi\rangle\,,\\
g_{\mu}(x)&=x_{\mu}\left[\frac{1}{2\pi^2x^4}-\frac{m^2}{8\pi^2x^2}+\frac{m}{16N}\langle
\overline{\psi}\psi\rangle\right].
\end{align}
Note that some one-loop corrections to this expression can be found, e.g.,
in Ref.~\cite{Gimenez:2005nt}.

We are interested in correlation functions of the type
\begin{align}\nonumber
\left\langle J^{(12)}(x)\overline{J}^{(12)}(0)\right\rangle
&=\pm\left\langle [\overline{\psi}_1\Gamma\psi_2](x)[\overline{\psi}_2\Gamma\psi_1](0)\right\rangle\\
&=\mp\left\langle\tr\left[\Gamma S_{F,2}(x)\Gamma\gamma_5S_{F,1}(x)\gamma_5
\right]\right\rangle,
\end{align}
where $S_{F,j}$ is the propagator of a quark of flavour $j$ and we used
the $\gamma_5$-Hermiticity $S_F^{\dagger}(0,x)=\gamma_5S_F(x,0)\gamma_5$.
The upper signs refer to $J\in\{S,P,V\}$ and the lower signs
to $J=A$.\footnote{These signs follow from the
convention Eq.~\eqref{eq:conve}. Different (pseudo)-Euclidean
conventions may result in
different signs.} Note that as we restrict ourselves to non-singlet
currents, the Wick contraction yields only one term.

It is now easy to see that
\begin{align}
\left\langle J^{(12)}(x)\overline{J}^{(12)}(0)\right\rangle
&=\mp N\left[
f_1(x)f_2(x)\tr(\Gamma^2\gamma_5^2)\right.
\nonumber\\&\quad+\left.g_{1\mu}(x)g_{2\nu}(x)\tr\left(\Gamma\gamma_{\nu}
\Gamma\gamma_5\gamma_{\mu}\gamma_5\right)\right]\nonumber\\
&=N\left[-4f_1(x)f_2(x)\right.\nonumber\\&\quad\pm \left.g_{1\mu}(x)g_{2\nu}(x)\tr\left(\Gamma\gamma_{\nu}\Gamma\gamma_{\mu}\right)\right].
\end{align}
Evaluating the above traces for the
combinations Eqs.~\eqref{eq corr s 12}--\eqref{eq corr a 12}
gives
\begin{align}
G_{J^{(12)}}(x)&=4N\left[-f_1(x)f_2(x)+s_{J}g_{1\mu}(x)g_{2\mu}(x)\right]\cdots \nonumber \\
&=\frac{N}{\pi^4}\frac{s_J}{x^6}
-\frac{N}{4\pi^4}\frac{m_1m_2+s_J(m_1^2+m_2^2)}{x^4}+\nonumber \\
&\quad+\frac{N}{16\pi^4}\frac{s_Jm_1^2m_2^2}{x^2}\nonumber\\
&\quad+
\frac{1}{8\pi^2}\frac{(2+s_J)(m_1+m_2)\langle\overline{\psi}\psi\rangle}{x^2}\nonumber \\
&\quad+\frac{1}{32\pi^2}\frac{s_J\langle FF\rangle}{x^2}+\cdots,
\label{eq:gggg}
\end{align}
where
\begin{align}
&s_S=1\,, &s_P=-1\,,\\
&s_V=-\frac12\,, &s_A=\frac12\,.
\end{align}
The contribution from the non-perturbative gluon condensate
that we added to Eq.~\eqref{eq:gggg}
is due to the possibility of a gluon coupling to each of
the quark lines and can be inferred from the results
of Refs.~\cite{Reinders:1981sy,Reinders:1984sr,Jamin:1992se}.
Note that up to the overall sign convention and
our prefactor $1/4$ in the definitions
Eqs.~\eqref{eq corr v 12} and \eqref{eq corr a 12} of $G_V$ and $G_A$,
the above result is consistent with the equal mass expressions obtained in
Ref.~\cite{Tomii:2016xiv}.
Four-loop radiative corrections for the massless case can be found
in Ref.~\cite{Chetyrkin:2010dx}.

Taking ratios of correlation functions obtained for different mass
parameters gives
\begin{align}
\frac{G_{J^{(12)}}(x)}
{G_{J^{(34)}}(x)}&=1+(A^J_{12}-A^J_{34})x^2\nonumber
\\&+\left[\left(A^J_{34}\right)^2-A^J_{12}A^J_{34} +B^J_{12}-B^J_{34}\right]x^4+\cdots,
\label{eq.rat}
\end{align}
with the mass dependent coefficients
\begin{align}
A^J_{jk}&=-\frac14\left(m_j^2+m_k^2+\frac{m_jm_k}{s_J}\right)\,,\label{eq.aa}\\
B^J_{jk}&=\frac{\pi^2}{32N}\langle FF\rangle+\frac{m_j^2m_k^2}{16}\nonumber\\
&\quad+\frac{\pi^2}{8N}\frac{2+s_J}{s_J}(m_j+m_k)\langle\overline{\psi}\psi\rangle\,.\label{eq.bb}
\end{align}
We will make use of this expression where, in our regime of quark masses,
the last term is the dominant one. The Gell-Mann--Oakes--Renner
relation
\begin{equation}
(m_j+m_k)\langle\overline{\psi}\psi\rangle=-F_0^2M_{jk}^2
\label{eq.gmor}
\end{equation}
can be used to substitute the chiral condensate term, 
thereby eliminating any free parameter.
$M_{jk}$ above denotes the mass of a pseudoscalar meson composed of
(anti)quarks of masses $m_j$ and $m_k$ and the
pion decay constant in the $N_f=3$ chiral limit
reads $F_0=86.5(1.2)\,\textmd{MeV}$~\cite{Aoki:2016frl,Agashe:2014kda}.
Note that
to order $x^4$ the gluon condensate does not contribute to
the ratio Eq.~\eqref{eq.rat} as it cancels
from the difference $B^J_{12}-B^J_{34}$.

\subsection{Lattice corrections}
\label{seclatc}
Now that we have worked out order $x^2$ and $x^4$ corrections,
we will also investigate the short distance, order $a$ corrections
to $b_J$.

The correlators can be
computed in lattice perturbation theory in a volume
of $N_t\times N^3$ sites. Unsurprisingly, we find
the result at short distances to depend only weakly
on the volume. Therefore, we employ antiperiodic fermionic
boundary conditions in time, in spite of the fact that
most of the analysed ensembles have open
boundaries~\cite{Luscher:2011kk,Bruno:2014jqa}.
We start from
the free Wilson quark propagator
\begin{equation}
S_F(x) = \frac{1}{(2\pi)^4}\sum_{p} \frac{-i\gamma_{\mu} \bar{p}_{\mu}
+ M(p)}{\sum_{\mu}\bar{p}_{\mu}^2 + M^2(p) } \exp\left(ipx\right)\,,
\end{equation}
where $\bar{p}_{\mu}=a^{-1}\sin(ap_{\mu})$, $p=(p_0,p_1,p_2,p_3)$ with
\begin{align}
p_0 &=-\frac{\pi}{a}+\frac{\pi}{N_ta}, -\frac{\pi}{a}+\frac{3\pi}{N_ta}, \ldots, \frac{\pi}{a}-\frac{\pi}{N_ta}, \nonumber \\  
p_i &= -\frac{\pi}{a}+\frac{2\pi}{Na}, \ldots, \frac{\pi}{a}\,,
\end{align}
and
\begin{equation}
M(p) = m_0 + \frac{2}{a} \sum_{\mu} \sin^2 \left( \frac{ap_{\mu}}{2} \right)\,.
\end{equation}
Using a simple computer program, we can evaluate and combine two of these
quark propagators into a correlator $G_J(x)$.
This then enables us to obtain (mass dependent) tree-level results
for the ratios $R_J$ and $\widetilde{R}_J$, see
Eqs.~\eqref{eq:rrat} and \eqref{eq:rrat2}.
We label these ratios as $R_J^{\mathrm{tree}}$ and $\widetilde{R}_J^{\mathrm{tree}}$.

Subtracting the tree-level expectation
from the lattice data
will not only reduce lattice artefacts but also the leading factor of
$1$ cancels identically from Eqs.~\eqref{eq:rrat} and \eqref{eq:rrat2}.
Moreover, the impact of the mass dependent perturbative $A^J_{jk}$
coefficients and of the $m_j^2m_k^2$ term within $B^J_{jk}$ (see
Eqs.~\eqref{eq.aa} and \eqref{eq.bb}) on Eq.~\eqref{eq.rat} is
removed to leading order. The effect of these terms was tiny
in any case in comparison to that of the chiral condensate appearing
within $B^J_{jk}$: $m\ll |\langle\overline{\psi}\psi\rangle|^{1/3}$.
Indeed, after subtracting
the leading order perturbative
expectation we are unable to resolve any remaining $x^2$
term within our numerical precision.

\begin{figure}
\begin{center}
\includegraphics[width=0.45\textwidth]{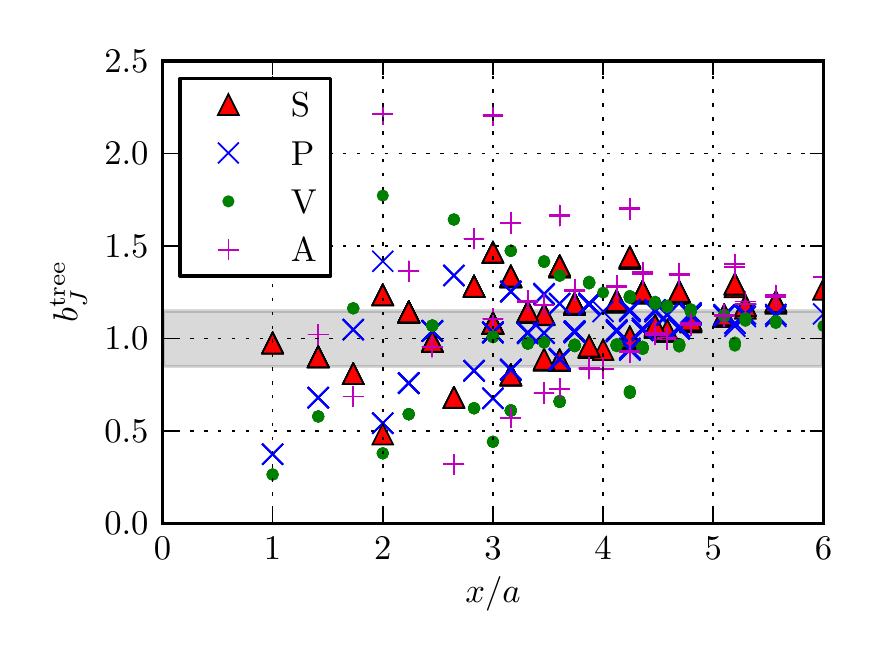}
\caption{Tree-level improvement coefficients. The data correspond 
to the largest mass difference that we can encounter.
The band indicates the region with cut-off effects smaller than 15\%. 
All separations $n=x/a$ within this band are accepted.\label{correction factors}}
\end{center}
\end{figure}

\subsection{Improved observables and the choice of the Euclidean distance}
Using the tree-level lattice perturbation theory results of
Sec.~\ref{seclatc} above as well as
Eqs.~\eqref{eq.rat}, \eqref{eq.bb} and \eqref{eq.gmor}, 
we define the improved ratio of correlators, cf.\ Eq.~\eqref{eq:rrat}:
\begin{widetext}
\begin{align}
B_J(x,m_s-m_{\ell})&\equiv
1+\biggl[R_J(x,m_s-m_{\ell})-R^{\mathrm{tree}}_J(x,m_s-m_{\ell})
\left.+\frac{\pi^2}{8N}\frac{2+s_J}{s_J}\left(M_{\pi}^2-M_K^2\right)F_0^2x^4\right]
\times\left(\frac{1}{\kappa_s}-\frac{1}{\kappa_{\ell}}\right)^{-1} \nonumber\\
&=b_J
+\mathcal{O}(x^6)+\mathcal{O}(g^2 a)+\mathcal{O}(g^2a^2/x^2)+\cdots\,,
\label{eq. obs bj}
\end{align}
where we have neglected small mass dependent terms of $\mathcal{O}(g^2x^2)$
and $\mathcal{O}(g^2x^4)$. $M_K=M_{13}$ and $M_{\pi}=M_{12}$ are the kaon
and pion masses, most of which are published
in Refs.~\cite{Bali:2016umi,Bruno:2014jqa}.
We also define $\widetilde{B}_J$, analogously generalizing Eq.~\eqref{eq:rrat2}:
\begin{align}
\widetilde{B}_J(x,\overline{m}^{(\sigma)}-\overline{m}^{(\rho)})&\equiv
1+\biggl[\widetilde{R}_J(x,\overline{m}^{(\sigma)}-\overline{m}^{(\rho)})
-\widetilde{R}^{\mathrm{tree}}_J(x,
\overline{m}^{(\sigma)}-\overline{m}^{(\rho)})
\left.+\frac{\pi^2}{8N}\frac{2+s_J}{s_J}\left({M_{\pi}^{(\rho)}}^2-
{M_{\pi}^{(\sigma)}}^2\right)F_0^2x^4\right]\nonumber\\
&\qquad\times\left(\frac{1}{\kappa^{(\sigma)}}-\frac{1}{\kappa^{(\rho)}}\right)^{-1}
=b_J+3\tilde{b}_J
+\mathcal{O}(x^6)+\mathcal{O}(g^2 a)+\mathcal{O}(g^2a^2/x^2)+\cdots\,.
\label{eq. obs bbj}
\end{align}
\end{widetext}
Before implementing the above equations, we must decide on the
$x=na$ distances to be considered. Differences between improvement
coefficients determined for different choices
of $x$ will be of order $a$, as long as $x^2\ll 1/\Lambda^2$.
In the end we will select one and the same lattice direction
$N_0$ to define our improvement
condition at $x_0\propto N_0a$. Ideally, for this choice order $a$
corrections to $b_J$ and to $\tilde{b}_J$ should be as small as possible.
As can be seen from the last term of Eqs.~\eqref{eq. obs bj} and
\eqref{eq. obs bbj}, $B_J$ and $\widetilde{B}_J$ need to be determined
at a fixed physical distance $|x|=|x_0|$. On a discrete
lattice $x_0$ cannot be kept fixed when changing the spacing,
however, we will use the $n_0\propto N_0$ value that
is closest to our choice of $x_0\approx n_0a$.
In addition to this reference point $n_0$, we realize
additional vectors $n$ to enable an estimation of the size of
$x^6$ and higher order continuum effects that have not been accounted for.

\begin{table}
\caption{Set of points selected for the non-perturbative analysis.
We average over equivalent directions in terms of the spatial
cubic symmetry including inversions and time reflections.
The point $N_0$ appears in boldface.
\label{table cutoff effects}}
\begin{center}
\begin{ruledtabular}
\begin{tabular}{cp{.4\textwidth}}
Channel &\multicolumn{1}{c}{$(n_0n_1n_2n_3)$}\\
\hline
$S$ & 
$(0 0 0 1)$,
$(0 0 1 1)$,
$(0 0 1 2)$,
$(0 1 1 2)$,
$(0 1 1 3)$,
${\bf(0 1 2 2)}$,
$(0 2 2 2)$,
$(1 0 0 0)$,
$(1 0 0 1)$,
$(1 0 0 2)$,
$(1 0 1 2)$,
$(1 0 1 3)$,
$(1 0 2 2)$,
$(1 1 2 3)$,
$(1 1 3 3)$,
$(1 2 2 3)$,
$(1 2 3 3)$,
$(2 0 0 1)$,
$(2 0 1 1)$,
$(2 0 1 2)$,
$(2 0 2 2)$,
$(2 1 1 3)$,
$(2 1 2 3)$,
$(2 1 3 3)$,
$(2 2 2 2)$,
$(2 2 2 3)$,
$(2 2 3 3)$,
$(3 0 1 1)$,
$(3 1 1 2)$,
$(3 1 1 3)$,
$(3 1 2 2)$,
$(3 1 2 3)$,
$(3 2 2 2)$\\
$P$ & 
$(0 1 1 1)$,
$(0 1 1 2)$,
$(0 1 1 3)$,
${\bf(0 1 2 2)}$,
$(0 1 2 3)$,
$(0 1 3 3)$,
$(0 2 2 2)$,
$(0 2 2 3)$,
$(0 2 3 3)$,
$(0 3 3 3)$,
$(1 0 1 1)$,
$(1 0 1 2)$,
$(1 0 1 3)$,
$(1 0 2 2)$,
$(1 0 2 3)$,
$(1 0 3 3)$,
$(2 0 1 1)$,
$(2 0 1 2)$,
$(2 0 1 3)$,
$(2 0 2 2)$,
$(2 0 2 3)$,
$(2 0 3 3)$,
$(3 0 0 3)$,
$(3 0 1 1)$,
$(3 0 1 2)$,
$(3 0 1 3)$,
$(3 0 2 2)$,
$(3 0 2 3)$\\
$V$ & 
$(0 1 1 2)$,
$(0 1 1 3)$,
${\bf(0 1 2 2)}$,
$(0 1 2 3)$,
$(0 1 3 3)$,
$(0 2 2 2)$,
$(0 2 2 3)$,
$(0 2 3 3)$,
$(0 3 3 3)$,
$(1 0 1 2)$,
$(1 0 1 3)$,
$(1 0 2 2)$,
$(1 0 2 3)$,
$(1 0 3 3)$,
$(2 0 1 1)$,
$(2 0 1 2)$,
$(2 0 1 3)$,
$(2 0 2 2)$,
$(2 0 2 3)$,
$(2 0 3 3)$,
$(3 0 1 1)$,
$(3 0 1 2)$,
$(3 0 1 3)$,
$(3 0 2 2)$,
$(3 0 2 3)$,
$(3 0 3 3)$\\
$A$ & 
$(0 0 1 1)$,
$(0 1 1 2)$,
${\bf(0 1 2 2)}$,
$(1 0 0 1)$,
$(1 0 1 2)$,
$(1 0 2 2)$,
$(1 1 3 3)$,
$(1 2 3 3)$,
$(2 0 1 1)$,
$(2 0 1 2)$,
$(2 1 3 3)$,
$(2 2 3 3)$,
$(3 1 1 3)$,
$(3 1 2 3)$,
$(3 2 2 3)$
\end{tabular}
\end{ruledtabular}
\end{center}
\end{table}

As a first step we define a subset of vectors for which tree-level
cut-off effects are small. In Fig.~\ref{correction factors} we show
improvement coefficients $b_J^{\textrm{tree}}$ evaluated in tree-level
lattice perturbation theory. For this comparison we set
$m_{\ell} = 0$ and $m_s$ equal to the
strange quark AWI mass obtained on the ensemble
H106, see Ref.~\cite{Bali:2016umi}. This mass approximately
corresponds to the physical strange quark mass,
obtained on the coarsest lattice spacing in use.
This choice represents the largest $a(m_s-m_{\ell})$ difference
that we can encounter. Note that in tree-level perturbation
theory the AWI and lattice quark masses coincide. The
higher order differences will be addressed in the discussion
of systematic errors, see Sec.~\ref{sec. condensates}.
For the subsequent analysis we select only those vectors $n$ for which
the deviation of $b_J^{\textrm{tree}}$ from the continuum expectation
$b_J=1$ is smaller than 15\%.

Table~\ref{table cutoff effects} summarizes the accepted
lattice vectors for the different currents.
One separation that is common to all the investigated channels
is $N_0=(0,1,2,2)$. For our present range of lattice
spacings, see Table~\ref{ensemble table},
this means $0.26\,\textmd{fm}>|N_0|a\geq 0.15\,\textmd{fm}$. We wish to
keep $|x_0|$ as small as possible to minimize the systematics.
A suitable compromise in view of the analysed lattice spacings is
$|x_0|=0.2\,\textmd{fm}$. Then, within this range $n_0=N_0$.
For $a<0.045\,\textmd{fm}$, which is the case for a set of
newly generated CLS ensembles at $\beta=3.85$, we will
have to increase $n_0=2N_0$ and for $a<0.027\,\textmd{fm}$
$n_0=3N_0$ to keep $|n_0|a\approx |x_0|$. Within the
range of lattice spacings that we cover here
$|x_0|$ varies by $\pm 25\%$ around the target value and
one may wonder about any associated systematics. We
investigate this in the Appendix.

We use the vector $n_0=N_0$ (and all its 24 equivalent permutations of
spatial components and signs) to compute the improvement coefficients.
All the remaining vectors within the range $0.15\,\textmd{fm}\lesssim
|n|a\lesssim 0.4\,\textmd{fm}$ are used to estimate systematic errors.

\section{Estimation of systematic errors}
\label{sec. systematics}
\subsection{Finite volume effects}
The improvement coefficients describe
short distance effects and therefore should be insensitive to the 
simulated volume. However, $b_J$ and $\tilde{b}_J$ obtained
using Eqs.~\eqref{eq. obs bj} and \eqref{eq. obs bbj}
will inherit the dependence of the correlation functions
$G_J$, that enter $R_J$ and $\widetilde{R}_J$, on $L=Na$.
For sufficiently small separations $|x|\ll L$ this dependence should
become negligible. Indeed, in a quenched setup
finite volume effects on ratios of massless
correlators evaluated at separations $|x|/L<1/8$ were found
to be below 1\% \cite{Cichy:2016qpu}. 
Extrapolating tree-level finite volume lattice perturbation theory
results to the continuum limit, keeping
$|x|/L=|n|/N$ and $mx$ fixed, and comparing the outcome to infinite
volume continuum perturbation theory, we confirmed that this
statement remains valid also for the ratio $R_J$ of Eq.~\eqref{eq:rrat}:
For $|x|/L < 1/8$ the differences are negligible, compared to
the other systematic errors that we will account for below.
We remark that all CLS ensembles satisfy the condition $LM_{\pi}\gtrsim 4$,
and hence the inequality $L>8|x_0|= 1.6\,\textmd{fm}$ always holds.
The worst case that we encounter corresponds to our coarse ``H''
ensembles where $|n_0|a\approx |x_0|$ takes its largest value in
physical units and $L/|n_0|a=32/3$.
Also note that in a non-perturbative setting finite size effects will
be even smaller, due to the mass gap.

\begin{figure*}
\begin{center}
\subfigure[Scalar]{\includegraphics[width=0.4\textwidth]{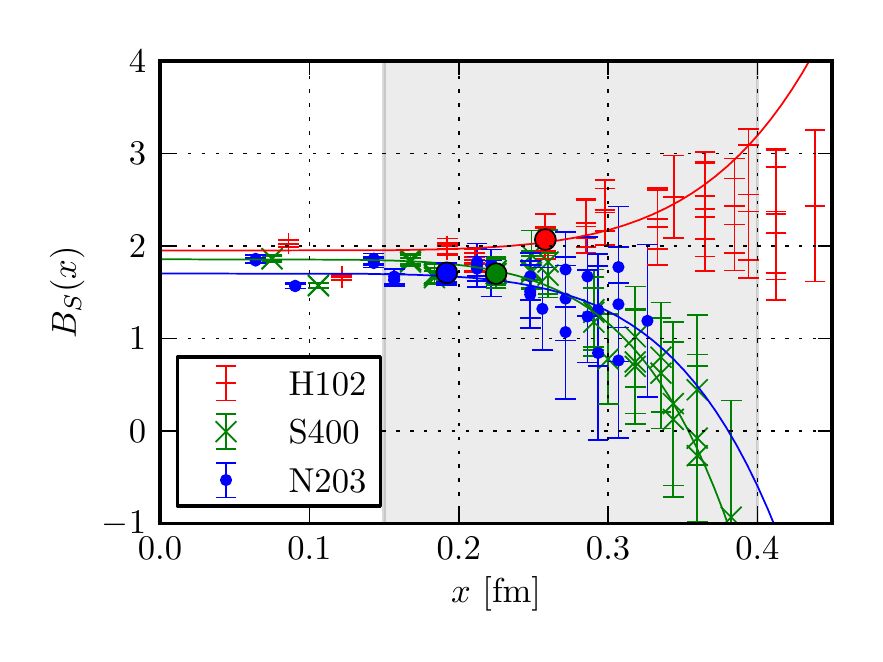}}
\subfigure[Pseudoscalar]{\includegraphics[width=0.4\textwidth]{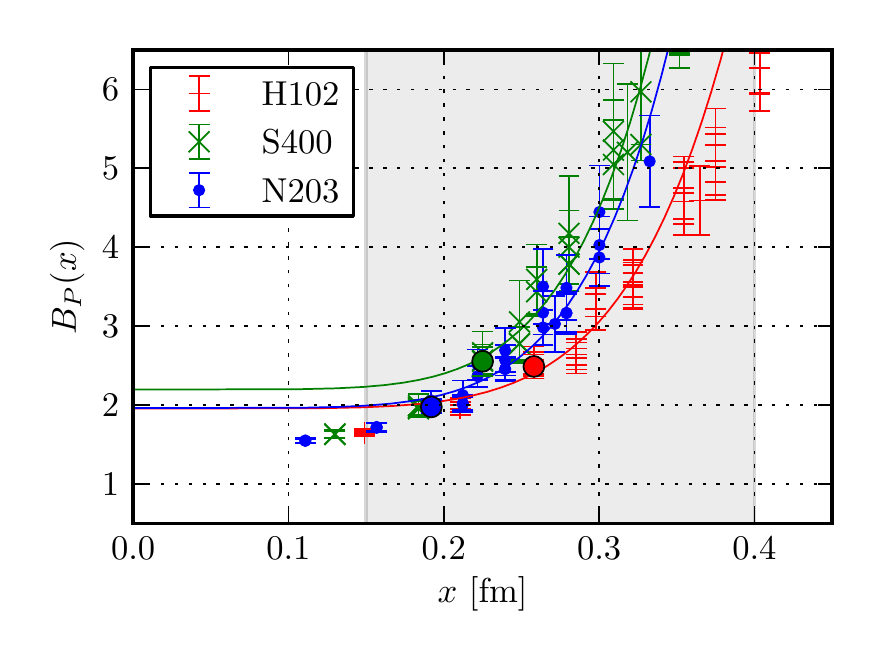}}
\subfigure[Vector]{\includegraphics[width=0.4\textwidth]{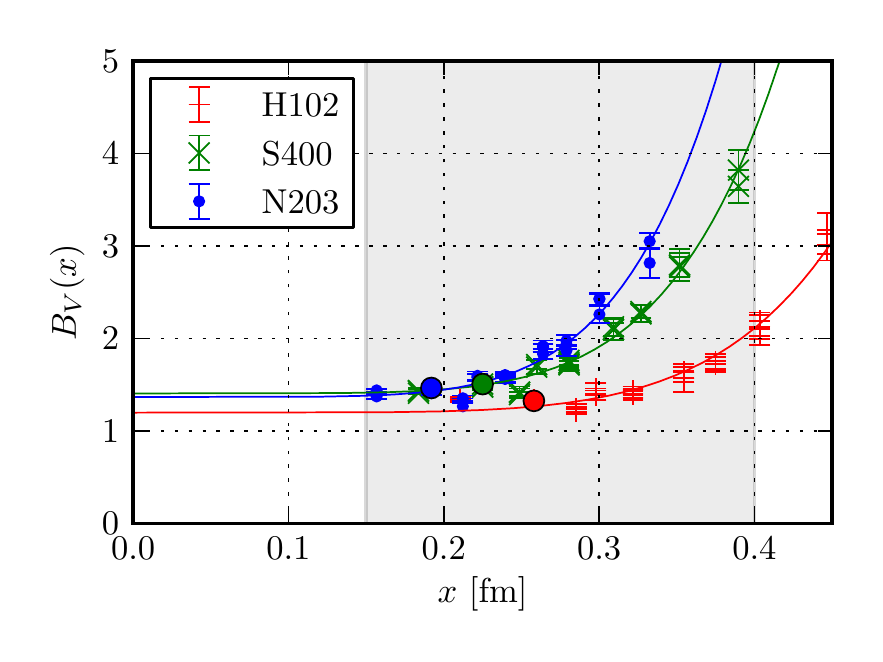}}
\subfigure[Axial]{\includegraphics[width=0.4\textwidth]{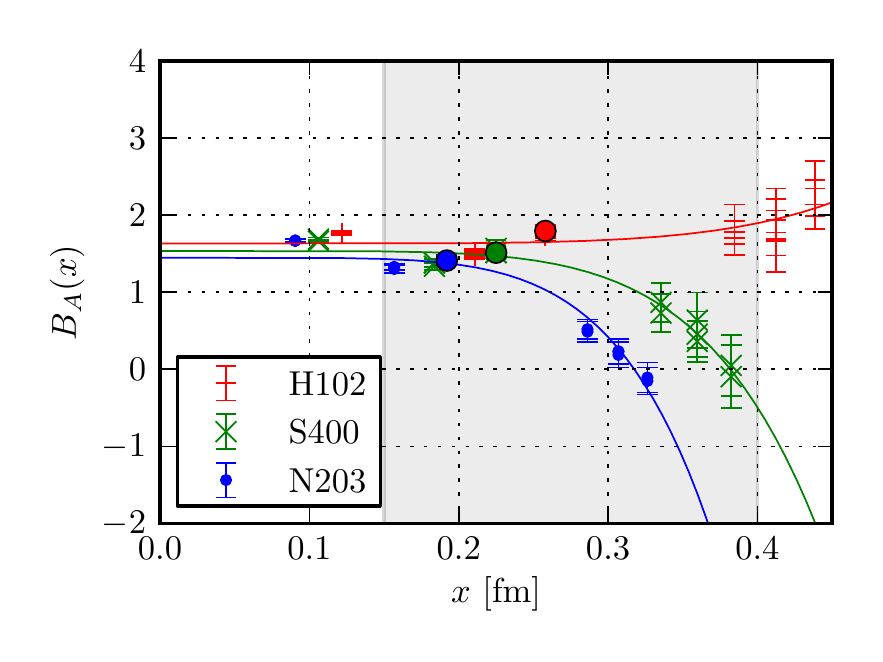}}
\caption{$B_{J}(x)$ for the ensembles H102, S400 and N203 which
share similar pion and kaon masses, see Table~\ref{ensemble table}. 
Solid lines denote the $x^6$ fit, Eq.~\eqref{eqx6}, to the data in the
shaded region. The $n_0a\approx x_0$ points that are used to 
define the $b_J$ coefficients are plotted as large circles.
The difference between a fit function at this position
from its value at $x=0$ constitutes one of our systematic errors.
\label{results bP}}
\end{center}
\end{figure*}

\subsection{Perturbative and non-perturbative corrections}
\label{sec. condensates}
Different conditions can be used to define the improvement
coefficients. As long as these definitions differ by
order $a$ terms all such schemes are equivalent in the sense
that improved expectation values of physical observables
will extrapolate to one and the same continuum limit,
linear in $a^2$, with $g^2a^2$, $a^3$ and higher order corrections.
For instance, we could have selected
a different value of $|x_0|$ along a different direction $N_0$:
Order $a$ corrections to
the $b_J$ and $\tilde{b}_J$ coefficients do not constitute a
source of systematics but correspond to a particular convention.

In Eqs.~\eqref{eq. obs bj} and \eqref{eq. obs bbj} we defined
the observables $B_J(x,m_s-m_{\ell})$ and
$\widetilde{B}_J(x,\overline{m}^{(\rho)}-\overline{m}^{(\sigma)})$, 
where the leading non-perturbative contribution (proportional to the
quark condensate) is subtracted and also the continuum perturbative
mass dependence is cancelled at tree-level. Higher order
mass dependent perturbative terms are neglected and
we employed AWI rather than renormalized quark masses, which will
also differ from each other at higher orders. For all
lattice spacings and quark mass combinations investigated,
at $n_0a\approx x_0$ we find these mass dependent tree-level
corrections to contribute only at the per mille
level to $B_J$ and $\widetilde{B}_J$.
Therefore, errors from neglecting the associated radiative corrections
will be completely insignificant in comparison to our statistical errors.
However, we also corrected
for the leading non-perturbative effect that is proportional to $x^4$.
This correction relies not only on the validity of the
Gell-Mann--Oakes--Renner relation in our regime of quark masses but
there exist also perturbative corrections to the Wilson coefficient,
that we have neglected. Figure~\ref{results bP} demonstrates that,
with the exception of the pseudoscalar
channel and up the scattering at short distances between different
lattice points, $B_J$ is almost perfectly flat up to distances
$|x|\approx 0.25\,\textmd{fm}$, indicating that such corrections
are small at $|x_0|=0.2\,\textmd{fm}$. We add 20\% of the subtracted
$x^4$ terms to our systematic error budget to reflect this
uncertainty. In order to discriminate whether the effect seen in the
pseudoscalar channel is due to particularly large radiative
corrections, lattice artefacts or interference from different
higher order terms, a perturbative calculation of the $x^4$ Wilson
coefficient is ongoing.

After adding the uncertainty of the $x^4$ subtraction to our error
budget, any remaining correction should be proportional to $x^6$
and higher orders. To account for this we fit 
\begin{equation}
\label{eqx6}
B_J(x) = \alpha_J + \beta_J x^6
\end{equation}
for each quark mass combination and current within the window
$0.15\,\textmd{fm}\lesssim|x|\lesssim0.4\,\textmd{fm}$.
A subset of these fits is shown in Fig.~\ref{results bP}. We quote
$|\beta_J (n_0a)^6|$  as a second systematic error. Note that these fits
are only performed to estimate the systematics and the curves shown do not
adequately represent the data, in particular at
short distances where lattice artefacts are visible and statistical
errors are small.

\begin{table*}
\caption{Summary of CLS (and RQCD) ensembles used in this study. ``O'' stands
for open and ``P'' for periodic boundary conditions (BC)
in time. $N_{\mathrm{conf}}$ denotes the number of analysed configurations
and ``sep.'' is the separations between two successive measurements
in molecular dynamics units.\label{ensemble table}}
\begin{center}
\begin{ruledtabular}
\begin{tabular}{cccccccccccc}
$\beta$ &$a/\textmd{fm}$&Ensemble& BC & $N_t$ & $N$ & $\kappa_{\ell}$ & $\kappa_s$ & $M_{\pi}$&$M_K$ &$N_{\mathrm{conf}}$& sep.\\
\hline
3.4 & 0.085&H101  & O & 96 & 32 & 0.136759 & 0.136759    &422&422& 100 & 40\\
3.4 & 0.085&H102  & O & 96 & 32 & 0.136865 & 0.136549339 &356&442& 100 & 40\\
3.4 & 0.085&H105  & O & 96 & 32 & 0.136970 & 0.13634079  &282&567& 103 & 20\\
3.4 & 0.085&H106  & O & 96 & 32 & 0.137016 & 0.136148704 &272&519&  57 & 20 \\
3.4 & 0.085&H107  & O & 96 & 32 & 0.136946 & 0.136203165 &368&549&  49 & 20\\
3.4 & 0.085&C101  & O & 96 & 48 & 0.137030 & 0.136222041 &223&476&  59 & 40\\
3.4 & 0.085&C102  & O & 96 & 48 & 0.137051 & 0.136129063 &223&504&  48 & 40\\
3.4 & 0.085&rqcd017& P & 32 & 32 & 0.1368650 & 0.1368650 &238&238& 150 & 20 \\
3.4 & 0.085&rqcd021& P & 32 & 32 & 0.136813  & 0.136813  &340&340&  50 & 20 \\
\hline               
3.46 & 0.076&S400 & O &128 & 32 & 0.136984 & 0.136702387 &354&446& 83 & 40\\
\hline               
3.55 & 0.064& N203 & O &128 & 48 & 0.137080 & 0.136840284&345&441& 74 & 40\\
\hline               
3.7 & 0.050 & J303  & O &192 & 64 & 0.137123 & 0.1367546608&260&478&38 & 40\\
\end{tabular}
\end{ruledtabular}
\end{center}
\end{table*}
\begin{figure*}
\begin{center}
\subfigure[Scalar]{\includegraphics[width=0.45\textwidth]{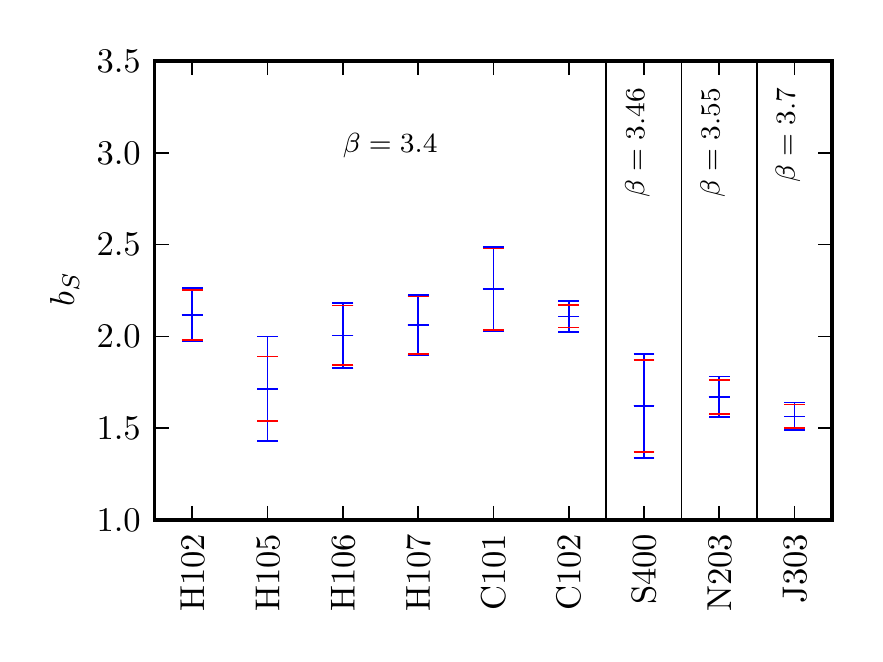}}
\subfigure[Pseudoscalar]{\includegraphics[width=0.45\textwidth]{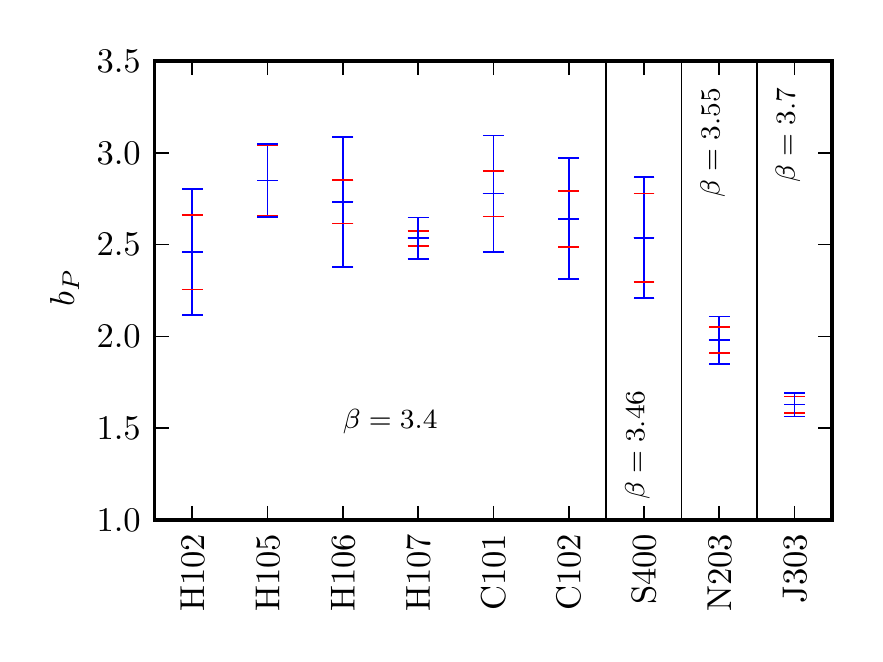}}
\subfigure[Vector]{\includegraphics[width=0.45\textwidth]{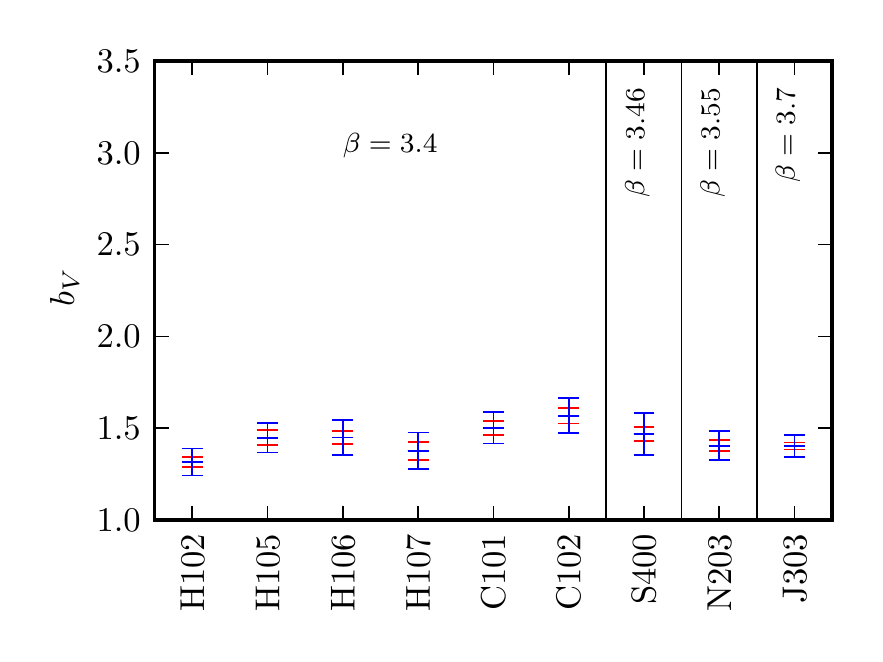}}
\subfigure[Axial]{\includegraphics[width=0.45\textwidth]{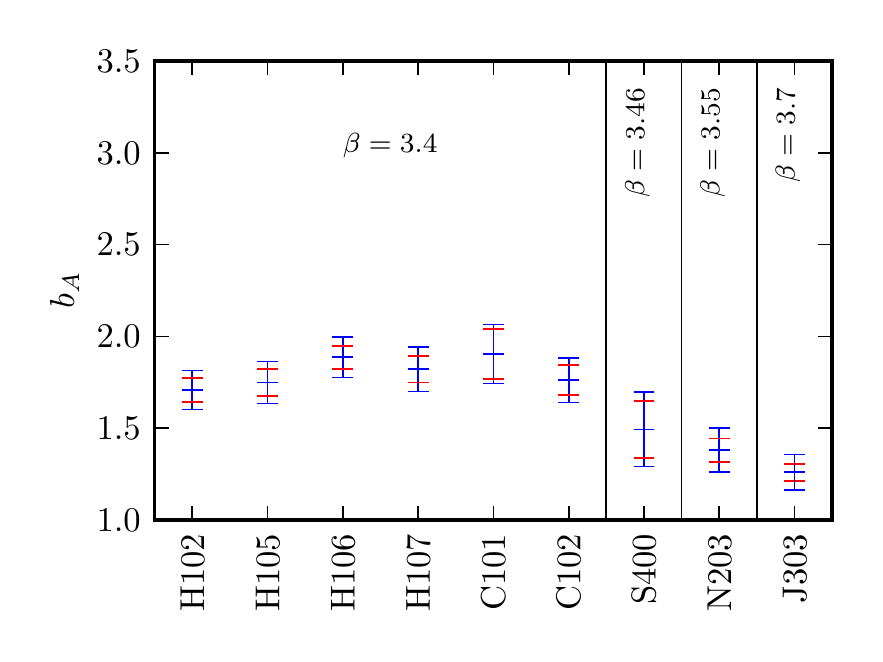}}
\caption{Numerical results for $b_J$ for all the mass non-degenerate ensembles.
The outer (blue) error bars denote the total uncertainties while the
inner (red) error bars indicate the statistical errors.\label{results final}}
\end{center}
\end{figure*}
\begin{table*}
\caption{Numerical results for the $b_{J}$ improvement coefficients.
The first error is statistical, the second error
corresponds to a 20\% uncertainty of the Wilson coefficient of the
non-perturbative $x^4$ correction and the third error is an estimate of
the size of order $x^6$ corrections, see
Sec.~\ref{sec. condensates}.\label{tab. b results}}
\begin{center}
\begin{ruledtabular}
\begin{tabular}{ccllll}
$\beta$ &Ensemble& $b_S$ & $b_P$ & $b_V$ & $b_A$ \\
\hline
3.4  & H102 & 2.12(14)(5)(1) & 2.46(20)(2)(28) & 1.38(3)(5)(5) & 1.71(7)(8)(2)\\
     & H105 & 1.72(18)(5)(22)& 2.85(19)(2)(50)  & 1.45(4)(5)(5) & 1.75(7)(8)(3)\\
     & H106 & 2.01(16)(5)(5) & 2.73(12)(2)(33) & 1.45(4)(5)(7) & 1.89(6)(9)(3)\\
     & H107 & 2.06(16)(5)(1) & 2.54(4)(2)(10)  & 1.38(5)(5)(7) & 1.82(7)(8)(5)\\
     & C101 & 2.26(23)(5)(1) & 2.78(13)(2)(29) & 1.50(4)(5)(6) & 1.90(14)(8)(3)\\
     & C102 & 2.11(6)(5)(3)  & 2.64(15)(2)(29) & 1.57(4)(5)(7) & 1.76(8)(8)(3)\\
\hline
3.46 & S400 & 1.62(22)(7)(11)& 2.54(24)(2)(23) & 1.46(4)(7)(8) & 1.49(15)(12)(6)\\
\hline
3.55 & N203 & 1.67(9)(5)(2)  & 1.98(7)(2)(11)  & 1.40(3)(5)(5) & 1.38(6)(9)(5)\\
\hline
3.7  & J303 & 1.56(6)(4)(1)  & 1.63(4)(1)(5)   & 1.40(2)(4)(4) & 1.26(5)(7)(5)
\end{tabular}
\end{ruledtabular}
\end{center}
\end{table*}
\begin{table}
\caption{Fit parameters for the parametrizations of
$b_{J}(g^2)$ Eqs.~\eqref{eqpade} and \eqref{eqpade2}
with the one-loop coefficients of Ref.~\cite{Taniguchi:1998pf}.
We also
include parametrizations for $\mathcal{A}=b_P-b_A+b_S$ and $b_A-b_P$.
Note that the leading order result for the latter combination
reads 0, instead of 1.
\label{tab. fit results}}
\begin{center}
\begin{ruledtabular}
\begin{tabular}{ccccc}
Coefficient&$b_J^{\text{one-loop}}$&$\gamma_J$&$\delta_J$&cov($\gamma_J$,$\delta_J$)\\
\hline
$b_S$&0.1526&$-0.439(50)$&$-0.535(14)$ & 0.972 \\
$b_P$&0.1187&$-0.354(54)$&$-0.540(11)$ & 0.945 \\
$b_V$&0.1181&$0.596(206)$& --- & --- \\
$b_A$&0.1175&$-0.523(33)$&$-0.554 (10)$& 0.984 \\
$\mathcal{A}$&0.1538&$-0.252(145)$&$-0.522(26)$& 0.973 \\
$b_A-b_P$&$-0.0012$&$22.6(20.7)$&$-0.512(62)$&0.968 
\end{tabular}
\end{ruledtabular}
\end{center}
\end{table}

\section{Results}
\label{sec. results}
We introduce the ensembles and analysis methods used,
before we present results on the $b_J$ coefficients, including an
interpolating parametrization, that is based on one-loop
perturbative results. For convenience we also
include two combinations of improvement
coefficients that are frequently needed.
Subsequently, we determine the $\tilde{b}_J$ coefficients
at our coarsest lattice spacing ($a\approx 0.085\,\textmd{fm}$,
$\beta=3.4$) as a proof of concept.
\subsection{Overview of the used ensembles}
We use ensembles generated within the
CLS initiative. In Table~\ref{ensemble table} we summarize
the ensembles employed and how many measurements
were taken; for more details see Refs.~\cite{Bruno:2014jqa,Bali:2016umi}.
Typically we perform 50 to 100 measurements, that are separated by
20 to 40 molecular dynamics units (MDUs) in the Markov Monte-Carlo chain.
The largest integrated autocorrelation time is associated with the
Wilson flow observable $E(t_0)$, which does not
exceed $\approx 100$~MDU, even at $\beta=3.7$~\cite{Bruno:2014jqa}.
Indeed, binning our data
gives no indication of autocorrelations.
The statistical errors are computed with the jackknife method.

\subsection{Results for \boldmath$b_{J}$}
\label{sec. results b}
\begin{figure*}
\begin{center}
\subfigure[Scalar]{\includegraphics[width=0.45\textwidth]{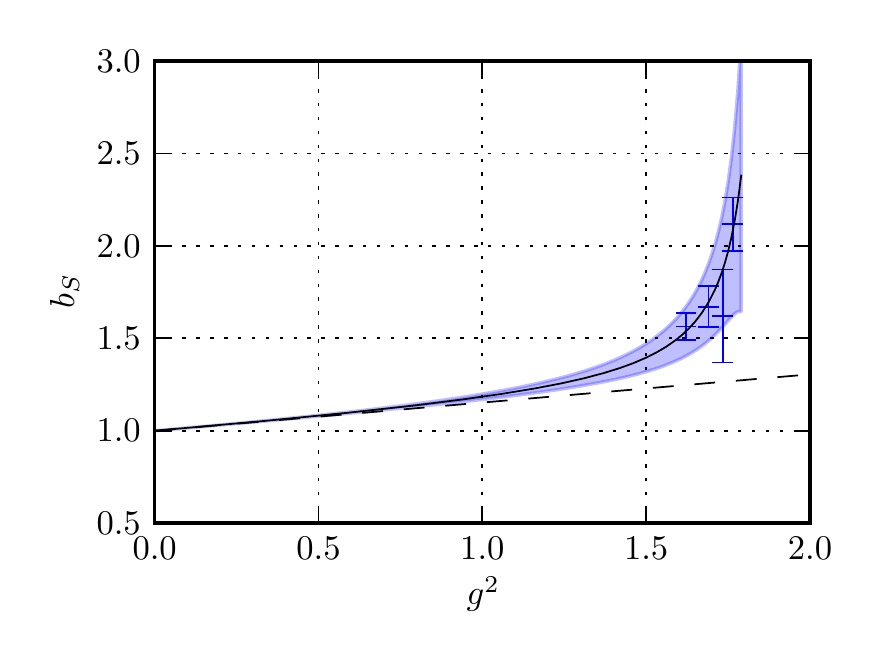}}
\subfigure[Pseudoscalar]{\includegraphics[width=0.45\textwidth]{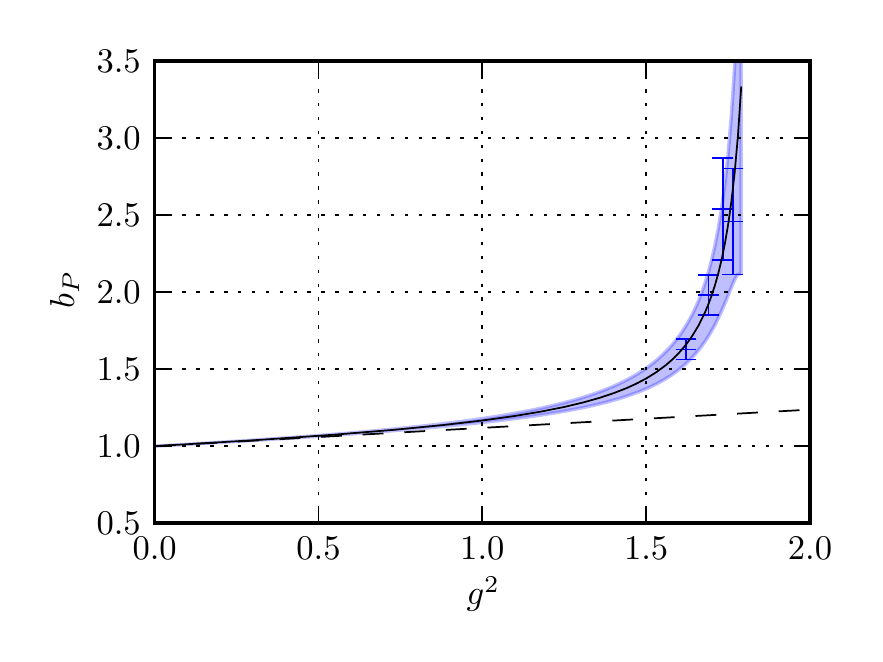}}
\subfigure[Vector]{\includegraphics[width=0.45\textwidth]{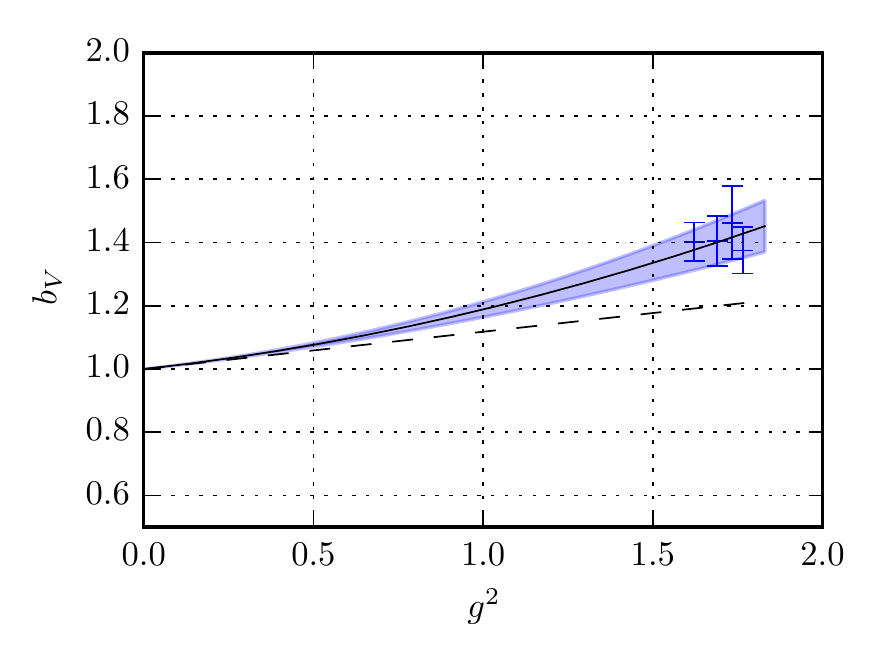}}
\subfigure[Axial]{\includegraphics[width=0.45\textwidth]{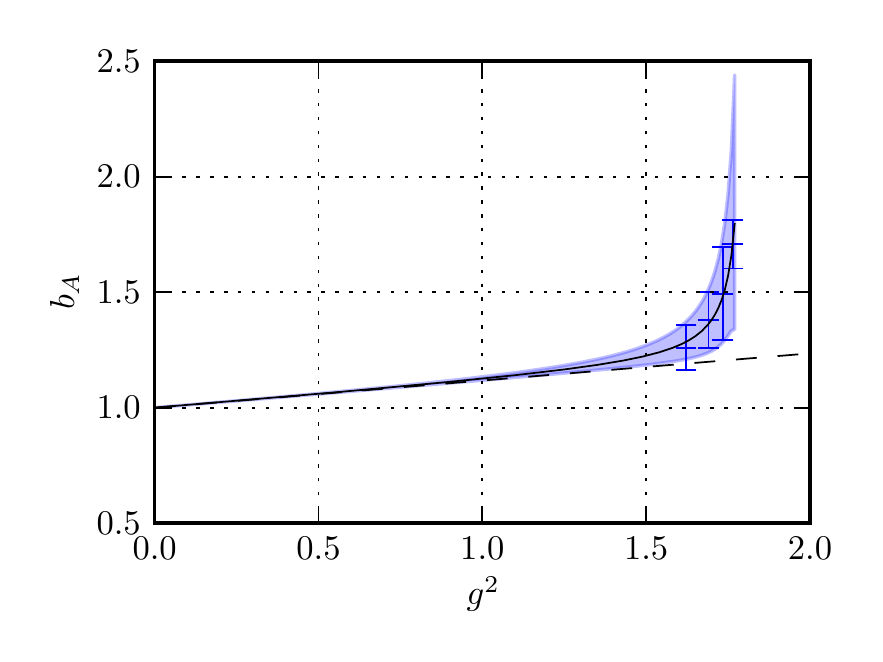}}
\caption{$b_J$ as functions of $g^2$. The error bands correspond to
Eqs.~\eqref{eqpade}--\eqref{eqpade2}
with the parameter values of Table~\ref{tab. fit results}. The dashed
lines are the perturbative one-loop expectations.
\label{pade}}
\end{center}
\end{figure*}

In Table~\ref{tab. b results} we list our results for each ensemble. 
These are also visualized in Fig.~\ref{results final}. Note that
at $\beta=3.4$ we have several mass combinations at our disposal,
giving several independent results that turn out to be compatible
with each other within
errors. We will quote the H102 results as our reference values since the pion
and kaon masses are similar in this case to those of S400 and N203.
Statistical and systematic errors depend on the lattice spacing and
quark mass combinations used and vary considerably between the ensembles.
Most of the results in the pseudoscalar channel
are dominated by the systematic uncertainties. However, the systematic errors
decrease as we approach finer lattices.
Note that the $b_J$ coefficients from the ensemble with the
finest lattice spacing, J303, have combined errors ranging from
4.8\% to 7.7\%, whereas
the relative uncertainty on $b_P$ determined on H102 amounts to 14\%.

We parameterize the $g^2=6/\beta$ dependence of our results
using a two parameter rational approximation:
\begin{align}
\label{eqpade}
b_J(g^2) &= 1 + b_J^{\textrm{one-loop}}g^2\frac{1+ \gamma_J g^2}{1+ \delta_J g^2}\quad \text{for}\quad J=S,P,A\,, \\
b_J(g^2) &= 1 + b_J^{\textrm{one-loop}}g^2 ( 1+ \gamma_J g^2 )\quad \text{for}\quad J=V\,.
\label{eqpade2}
\end{align}
Note that $b_V$ is well described by a one parameter fit, such that in this case
allowing for $\delta_V\neq 0$ does not result in a stable fit.
The parameters
\begin{align}
b_P^{\textrm{one-loop}} &= 0.0890(1) C_F\,, &b_S^{\textrm{one-loop}} = 0.11444(1) C_F\,, \\
b_A^{\textrm{one-loop}} &= 0.0881(1) C_F\,, &b_V^{\textrm{one-loop}} = 0.0886(2) C_F\,,
\label{eqoneloop}
\end{align}
where $C_F=4/3$,
correspond to the one-loop coefficients that were computed for our
action in Ref.~\cite{Taniguchi:1998pf},
so that the parametrizations respect the known perturbative limits.

We include the ensembles H102, S400, N203 and J303 into our fit.
Note that H102, S400 and N203
share similar pion and kaon masses. We combine
the statistical and systematic uncertainties in quadrature.
The fits are shown in Fig.~\ref{pade} and
the fit parameter values are collected in Table~\ref{tab. fit results}. 
The parameter values for the two parameter fits are highly
correlated and we give the correlation coefficients in the last column
of the table. This, along with the statistical errors
of $\gamma_J$ and $\delta_J$, is used to generate the error bands shown
in the figure. 
In the case of $\gamma_V$, where we carried out a one parameter fit,
we obtained a value $\chi^2/3=0.29$ and rescaled the error on the
fit parameter (and the error band shown) by $\sqrt{1/0.29}$ to be on
the safe side.
We remark that for our action no simulations are planned
for $g^2$ values outside the bands shown, i.e.\ for $g^2>1.8$, where
the above rational parametrizations exhibit poles.

One often encounters specific linear combinations of improvement
coefficients. In Refs.~\cite{Bhattacharya:2005rb} and
\cite{Bali:2016umi} the coefficient
\begin{equation}
\mathcal{A} = b_P - b_A + b_S=b_P-b_A-2b_m
\end{equation}
plays an important role
while the combination $b_A-b_P$ is needed to convert
AWI into renormalized quark masses.
Therefore, we specifically analyse these combinations too
and include the corresponding rational
parametrizations in the last two lines of Table~\ref{tab. fit results}.
Note that the small value of $b^{\text{one-loop}}_{A-P}=-0.0012$ for the
combination $b_A-b_P$ results in a large $\gamma_{A-P}$ coefficient.
This also means that for the parametrization to be accurate,
in this case it is important to set
$b^{\text{one-loop}}_{A-P}$ exactly to this value, ignoring its
uncertainty of approximately $2\cdot 10^{-4}$. 

\subsection{Results for \boldmath$\tilde{b}_{J}$}
\label{sec. results bar}
The $\tilde{b}_{J}$ improvement coefficients carry much
larger statistical errors than their $b_J$ counterparts
since one needs to combine data from at least two independent ensembles.
Therefore, the errors cannot benefit from correlations between
statistical fluctuations but
always add up.

First, we compute $\widetilde{B}_J$ for pairs of
the three symmetric
line ensembles rqcd021, rqcd017 and H101 and find consistent
results. Next, in order to combine information from all three
ensembles, we correct the individual $\widetilde{R}_J$ ratios 
defined in Eq.~\eqref{eq:rrat2} in the way suggested by
Eq.~\eqref{eq. obs bbj} and extract the combination
$b_J+3\tilde{b}_J$ from the linear slope
of the $1/\kappa$ dependence at $n_0a\approx x_0$.
Following the procedure outlined in Sec.~\ref{sec. condensates}, we
then allow for 20\% of the $x^4$ correction term as one systematic
error and add another error, associated to the $x^6$ term from a fit
according to
Eq.~\eqref{eqx6}.
Finally, we subtract the $b_J$ values obtained on ensemble H102, see
Table~\ref{tab. b results}, to
arrive at the results
\begin{align}
\label{eq:btil1}
\tilde{b}_S &=  \ \ 0.9 \ (5.4) \ (0.1) \ (0.9)\,, \\
\tilde{b}_P &=  -6.8 \ (4.5) \ (0.1) \ (0.2)\,, \\
\tilde{b}_V &=  -3.5 \ (2.8) \ (0.1) \ (0.4) \,,\\
\tilde{b}_A &= \ \  0.5 \ (3.4) \ (0.2) \ (0.3) \,, 
\label{eq:btil2}
\end{align}
where the first errors are statistical and the other two uncertainties
correspond to the systematic errors explained above. Within large
statistical errors, that dominate the error budget,
all values are consistent with zero.

\section{Conclusions and outlook}
\label{sec. conclusions}
We computed ``$b_J$'' improvement coefficients parameterizing the
linear cut-off effects that are proportional to non-singlet quark mass
combinations for flavour non-singlet quark bilinear currents
on a set of CLS ensembles at four lattice spacings:
$a\approx 0.085\,\textmd{fm}$,
$a\approx 0.076\,\textmd{fm}$, $a\approx 0.064\,\textmd{fm}$
and $a \approx 0.05\,\textmd{fm}$. We also provide first estimates
of the $\tilde{b}_J$ coefficients that accompany the trace of
the quark mass matrix.

Our method is based on the short distance
behaviour of current-current correlation functions and turned out
to be statistically very precise, given the relatively small
computational effort.
We benefited from subtracting the dominant non-perturbative effects
as well as the leading perturbative lattice artefacts.
We carefully investigated systematic errors related to
non-perturbative and perturbative corrections as well as
finite volume effects and included the relevant uncertainties
into the errors of the results that we quote.

Our main result is the parametrization of the $b_J$ coefficients
Eqs.~\eqref{eqpade} and \eqref{eqpade2} with the parameter
values given in Table~\ref{tab. fit results}, which is valid
for the range $3.4\leq \beta\leq 3.7$. In the future we
will extend this range towards higher $\beta$ values and
also increase the statistical precision. These coefficients
become very important for heavy quark masses like that of the
charm; for instance $b_A$ significantly contributes to charmed
pseudoscalar meson decay constants. Neither can their effect be neglected if
one is interested in matrix elements involving strange quarks
within a (sub) percent level accuracy.

Preliminary results at our coarsest lattice spacing were obtained for
the $\tilde{b}_J$ parameters too, see
Eqs.~\eqref{eq:btil1}--\eqref{eq:btil2}. In this case we had to combine
data from different gauge ensembles and could not benefit from cancellations
of statistical fluctuations.
This means that more measurements are required.
Since the $\tilde{b}_J$ originate from sea quark effects, these
are of order $g^4$ in perturbation theory. However, within our
present uncertainties we cannot exclude large values of these coefficients,
and our preliminary results in fact suggest that some of them
may be unusually large.

It is known that the ratio of the singlet over the non-singlet mass
renormalization constant $r_m$ is about $2.6$ at $\beta=3.4$ and
still $1.5$ at $\beta=3.55$~\cite{Bali:2016umi}, far from the asymptotic
value of one. As a consequence of this decrease of $r_m$ with $\beta$,
starting from a relatively high value, the combination
$(2m_{\ell}+m_s)a=(2/\kappa_{\ell}+1/\kappa_s-3/\kappa_{\mathrm{crit}})/2$
stays fairly constant within the range of investigated lattice spacings
at fixed renormalized quark mass values, while naively one would
have expected it to decrease with $a$. The sea always contains the
relatively heavy strange quark, so that at realistic values
of the sea quark masses
the above combination (that accompanies $\tilde{b}_J$) is about
0.012 and 0.014~\cite{Bali:2016umi} at $\beta=3.4$ and
$\beta=3.55$, respectively. Therefore, a
value $\tilde{b}_A=1$ would increase light pseudoscalar decay constants
by more than 1\%. Clearly, this needs to be investigated further,
including $\beta>3.4$ and significantly increasing statistics,
to enable a full order $a$ improved continuum limit to be taken for
a wide range of physical observables. We also plan to extend the present
study to different currents, including flavour-singlet operators.

\begin{acknowledgments}
We used the {\sc CHROMA}~\cite{Edwards:2004sx} software package along with the
the multigrid solver implementation of Ref.~\cite{Heybrock:2015kpy} (see also 
Refs.~\cite{Heybrock:2014iga, Frommer:2013fsa}). Computations were performed on Regensburg's QPACE B
Xeon Phi system and on the SFB/TRR 55 QPACE 2 system \cite{Arts:2015jia} hosted
in Regensburg. We thank Wolfgang S\"oldner, Peter Georg,
Benjamin Gl\"a\ss le and Daniel Richtmann for their help
and technical support. We thank Antonio Pineda for discovering
relevant literature and Vladimir Braun for discussions as well as
all our CLS colleagues. We also thank Tassos Vladikas for a very careful reading of the manuscript.
\end{acknowledgments}
\appendix
\section{Impact of the quantization of lattice distances}
The distance $|x_0|$ used to determine the improvement coefficients
can in principle be chosen at will as long as $|x_0|=|n_0|a$ is
kept (approximately) constant, to achieve the complete
removal of order $a$ terms from continuum limit extrapolations of
physical observables. We restricted ourselves to points
where lattice artefacts on the $b_J$ coefficients are small --- at least at
tree-level. Keeping $|n_0|$ constant (rather than $|x_0|$)
would result in corrections to the $b_J$ coefficients of order $1/|n_0|^2$
(rather than of order $a^2/x_0^2$), which will not vanish as the continuum
limit is taken. We rely on non-perturbative corrections to be small in the
continuum theory, which means $|x_0|\lesssim 0.25\,\textmd{fm}$ is a
necessary condition for the method to be applicable. On the coarsest
lattices of interest this translates into $|x_0|\lesssim 3a$. Clearly,
at such separations the quantization of lattice distances cannot
be neglected and indeed the data points shown in Fig.~\ref{results bP}
at very short distances are not well described by continuous curves.

In this article we decided to take $n_0=mN_0$ along a fixed
lattice direction $N_0=(0,1,2,2)$, which corresponds to the smallest
distance appearing within all four channels shown in
Table~\ref{table cutoff effects}. We then set the
multiple $m\in\mathbb{N}$ such that $m|N_0|a$ was closest to
$|x_0|=0.2\,\textmd{fm}$. In our case this meant $m=1$ for all four lattice
spacings, and we investigate the systematics of this approximation in this
Appendix.

\begin{table}
\caption{The lattice points of Table~\protect\ref{table cutoff effects}
that are closest to $|x_0|=0.2\,\textmd{fm}$
from below ($x_{0<}$) and from above ($x_{0>}$) for the channels
$J\in\{S,P,V,A\}$. In the case of $\beta=3.4$ the point $x_{0<}$ does not
exist for $J=A$. Whenever $x_{0<}=N_0a$ or $x_{0>}=N_0a$ this is indicated
using boldface.
\label{tab:lattpoints}}
\begin{center}
\begin{ruledtabular}
\begin{tabular}{cccccc}
$J$&$\beta$&$x_{0<}/a$&$x_{0>}/a$&$|x_{0<}|/\textmd{fm}$&$|x_{0>}|/\textmd{fm}$\\\hline
   &$3.4 $&$(0 0 1 2)$&$(0 1 1 2)$& 0.191& 0.209   \\
$S$&$3.46$&$(0 1 1 2)$&${\bf(0 1 2 2)}$& 0.186& 0.228   \\
   &$3.55$&${\bf(0 1 2 2)}$&$(0 1 1 3)$& 0.193& 0.214   \\
   &$3.7 $&$(1 1 2 3)$&$(1 2 2 3)$& 0.194& 0.212   \\\hline
   &$3.4 $&$(0 1 1 1)$&$(0 1 1 2)$& 0.148& 0.209   \\
$P$&$3.46$&$(0 1 1 2)$&${\bf(0 1 2 2)}$& 0.186& 0.228   \\
   &$3.55$&${\bf(0 1 2 2)}$&$(0 1 1 3)$& 0.193& 0.214   \\
   &$3.7 $&$(0 1 2 3)$&$(0 2 2 3)$& 0.187& 0.206   \\\hline
   &$3.4 $& ---       &$(0 1 1 2)$&  --- & 0.209   \\
$V$&$3.46$&$(0 1 1 2)$&${\bf(0 1 2 2)}$& 0.186& 0.228   \\
   &$3.55$&${\bf(0 1 2 2)}$&$(0 1 1 3)$& 0.193& 0.214   \\
   &$3.7 $&$(0 1 2 3)$&$(0 2 2 3)$& 0.187& 0.206   \\\hline
   &$3.4 $&$(0 0 1 1)$&$(0 1 1 2)$& 0.121& 0.209   \\
$A$&$3.46$&$(0 1 1 2)$&${\bf(0 1 2 2)}$& 0.186& 0.228   \\
   &$3.55$&${\bf(0 1 2 2)}$&$(0 1 1 3)$& 0.193& 0.214   \\
   &$3.7 $&${\bf(0 1 2 2)}$&$(1 1 3 3)$& 0.150& 0.224
\end{tabular}
\end{ruledtabular}
\end{center}
\end{table}

\begin{figure}
\begin{center}
\includegraphics[width=0.49\textwidth]{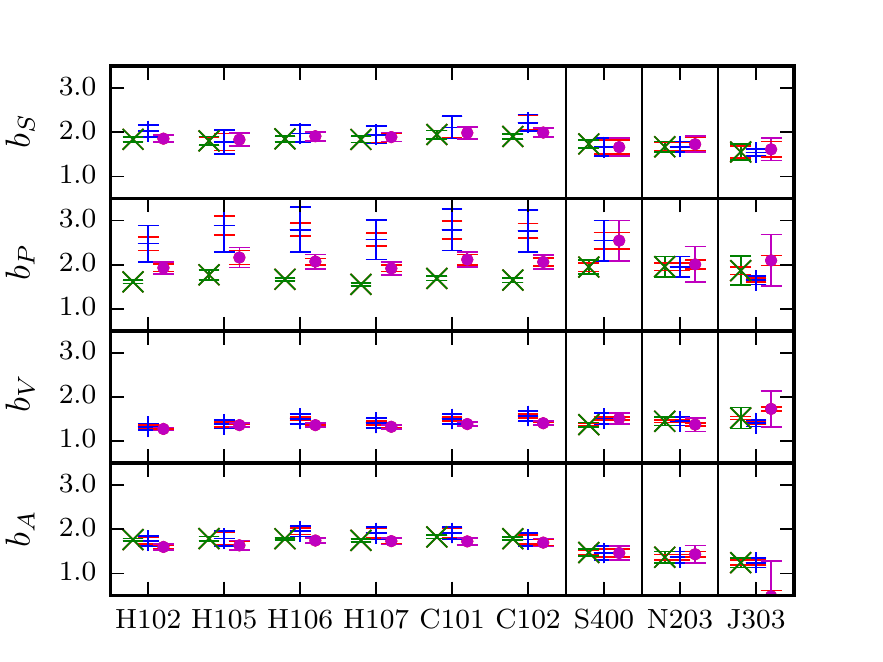}
\caption{The improvement coefficients $b_J$,
extracted for different choices of $x_0$. In addition to
$x_0=N_0a$ (blue central point within each group of three points),
the results obtained for $x_{0<}$ (left points where available) and
$x_{0>}$ (right points) are shown, see Table~\protect\ref{tab:lattpoints}.
Inner error bars are statistical only. Outer error bars include systematics.
The ``H'' and ``C'' ensembles correspond to $\beta=3.4$ ($a\approx 0.085\,\textmd{fm}$), S400 to
$\beta=3.46$, N203 to $\beta=3.55$ and J303 to $\beta=3.7$
($a\approx 0.050\,\textmd{fm}$).\label{fig:testdist}}
\end{center}
\end{figure}

In Table~\ref{tab:lattpoints} we list for our four
currents and four $\beta$ values the lattice distances
$|x_{0<}|$ and $|x_{0>}|$ that are closest to $|x_0|=0.2\,\textmd{fm}$
from below and from above, within the set of points of small tree-level
artefacts listed in Table~\ref{table cutoff effects}.
This is to be compared to the $|n_0|a=|N_0|a$ values of
$0.256\,\textmd{fm}$,
$0.228\,\textmd{fm}$,
$0.193\,\textmd{fm}$ and
$0.150\,\textmd{fm}$ at $\beta=3.4$, $3.46$, $3.55$ and
$3.7$, respectively.
We could have relaxed the restriction to one lattice direction $N_0$
and for instance have used the $|x_{0<}|$ or $|x_{0>}|$ values
(or an average of these) to define the improvement coefficients. 
In Fig.~\ref{fig:testdist} we compare the results of such different
strategies. For each group of three points the central point corresponds
to the result obtained using $N_0a$ with the point on the left
corresponding to $x_{0<}$ and to the right to $x_{0>}$. For the vector channel
at $\beta=3.4$ no $x_{0<}$ point exists and in some cases
either $x_{0<}$ or $x_{0>}$ happen to coincide with $N_0a$, see
Table~\ref{tab:lattpoints}. At $\beta=3.7$ even $|x_{0<}|$ is larger than
$|N_0|a$. Within present errors the different results mostly
appear to be consistent. The deviation of
$|N_0|a$ from $x_0=0.2\,\textmd{fm}$ is largest at $\beta=3.4$
and $\beta=3.7$ and at $\beta=3.4$ there appears to be
some tension in the pseudoscalar channel. This will be addressed with
increased precision in the near future.
\bibliography{references2}
\end{document}